\documentclass[12pt]{article}
\usepackage{amsmath,amssymb,epsfig,ulem}
\usepackage[bookmarks=false]{hyperref}
\usepackage{color}
\usepackage{multicol}

\allowdisplaybreaks



\newcommand{\beq}{\begin{equation}}
\newcommand{\eeq}{\end{equation}}
\newcommand{\bea}{\begin{eqnarray}} 
\newcommand{\eea}{\end{eqnarray}}

\begin{document}

\begin{titlepage}

\begin{flushright}
MITP/13-005\\
PITT-PACC-1304
\end{flushright}

\vspace{15pt}
\begin{center}
  \Large\bf Top-Quark Charge Asymmetry Goes Forward:\\
Two New Observables for Hadron Colliders
\end{center}

\vspace{5pt}
\begin{center}
{\sc Stefan Berge$^{a}$ and Susanne Westhoff$^{b}$}\\
\vspace{10pt} {\sl
$^{a}$ PRISMA Cluster of Excellence, Institut f\"ur Physik (WA THEP), \\
Johannes Gutenberg-Universit\"at,
D-55099 Mainz, Germany} \\
\vspace{10pt} {\sl
$^b$ PITTsburgh Particle physics, Astrophysics \& Cosmology Center (PITT PACC), \\
Department of Physics and Astronomy, University of Pittsburgh, Pittsburgh, PA 15260, USA} \\
\end{center}

\vspace{10pt}
\begin{abstract}
\vspace{2pt} 
\noindent
We propose two new observables to measure the charge asymmetry in hadronic top-quark pair production in association with a hard jet. The {\it incline asymmetry}, based on the inclination between the planes of initial- and final-state momenta, probes the charge asymmetry in the quark-antiquark channel. Compared to the hitherto investigated rapidity asymmetries, the incline asymmetry provides improved access to the partonic charge asymmetry at both the Tevatron and the LHC. The {\it energy asymmetry}, based on the energy difference between top and antitop quarks, for the first time allows us to probe the charge asymmetry in the quark-gluon channel at the LHC. In quantum {chro\-mo\-dy\-na\-mics}, asymmetries of up to $-12\,\%$ at the leading order are achievable with appropriate cuts. Top-pair plus jet production thus has the potential to become the discovery channel of the charge asymmetry in proton-proton collisions.
\end{abstract}

\end{titlepage}
\section{Introduction}
The top quark is often suspected to play a special role among the quarks of the standard model (SM). Its strong coupling to the Higgs boson suggests that the top quark may be involved in the mechanism of electroweak symmetry breaking and/or the generation of fermion masses. Thanks to the recent progress in top-quark physics on both the theoretical and the experimental side, those fundamental issues can now be addressed at an unprecedented level of precision. To date, top-quark observables at hadron colliders have passed the various tests of top-quark properties in the SM. One exception is the top-quark charge asymmetry, which has been measured as a forward-backward asymmetry in inclusive top-antitop production at the Tevatron \cite{Aaltonen:2012it,Abazov:2011rq}. It exceeds the SM prediction by about three standard deviations. The limited set of data recorded during the runtime of the Tevatron makes it difficult to draw a conclusion on the origin of this discrepancy. At the LHC, the charge asymmetry in top-quark pair production is buried under the large charge-symmetric background from partonic gluon-gluon ($gg$) initial states. Measurements of the charge asymmetry by the ATLAS and CMS collaborations do not indicate a deviation from the SM prediction, but they are affected by large uncertainties \cite{ATLAS:2012an,ATLAS:acll,Chatrchyan:2012cxa,CMS:acll}.

Novel information on the top quark's properties can be obtained by investigating top-quark pair production in association with a hard jet. Beyond the standard model, $t\bar t + j$ production provides a wide playground for sensitive probes of new physics, which recently started to be explored \cite{Gresham:2011dg,Knapen:2011hu,Ferrario:2009ee,Berge:2012rc}. In this work, we will focus on the charge asymmetry in quantum chromodynamics (QCD). The presence of an additional jet allows us to investigate QCD effects at leading order (LO), which contribute to inclusive top-pair production at next-to-leading order (NLO). In inclusive $t\bar t$ production, the charge asymmetry arises at NLO from virtual and real gluon radiation \cite{Kuhn:1998kw}. The same real gluon contributions generate the asymmetry in $t\bar t + j$ production at LO. While the completion of NNLO QCD calculations for the inclusive $t\bar t$ asymmetry is still ongoing, NLO corrections to the asymmetry in $t\bar t + j$ have been calculated and found to be sizeable \cite{Dittmaier:2007wz,Dittmaier:2008uj,Melnikov:2010iu}. The effects of top-quark decay and parton showers have been investigated in detail \cite{Melnikov:2011qx,Kardos:2011qa,Alioli:2011as}. The SM prediction of the charge asymmetry in $t\bar t + j$ thus has a solid basis, ready to face experimental results. At the Tevatron, the production rate of $t\bar t + j$ final states is relatively small due to the limited phase space \cite{CDF:ttjxs}. With higher collision energies at the LHC, $t\bar t$ events are produced abundantly in association with at least one hard jet \cite{ATLAS:ttjxs,CMS:ttjxs}. From a statistical point of view, the measurement of a charge asymmetry in $t\bar t + j$ production is therefore within reach.

A detailed understanding of the jet kinematics in $t\bar t + j$ production is crucial in order to control the infrared and collinear behavior of the process. As a result of the enhanced symmetric cross section in the collinear limit, the normalized charge asymmetry is maximized if the jet is emitted perpendicular to the beam axis in the partonic center-of-mass (CM) frame \cite{Berge:2012rc}. This feature facilitates the construction of a collinear-safe observable by setting an experimental cut on the jet rapidity in the parton frame. The jet handle can be further exploited to probe the charge asymmetry in $t\bar t + j$ production through observables that are not accessible in inclusive $t\bar t$ production. The construction and investigation of such observables is the purpose of this work.

We aim at providing optimal access to the charge asymmetry in QCD in both the partonic quark-antiquark ($q\bar q$) channel and the quark-gluon ($qg$) channel of $t\bar t + j$ production. To this end, we explore the relation between the momenta of the top quarks and the jet in the final state. We show that the angular correlation between the light quarks inside the proton and the top quarks in the final state is appropriately described by the inclination between the planes spanned by the light-quark and (gluon-)jet momenta and by the top-quark and jet momenta, respectively. The corresponding observable, called {\it incline asymmetry}, probes the charge asymmetry in the $q\bar q$ channel. It is superior to the previously studied forward-backward or rapidity asymmetries, which do not take the jet kinematics into account. The incline asymmetry can be formulated for both proton-antiproton collisions at the Tevatron and proton-proton collisions at the LHC.

Due to the relation between the particles' four-momenta in the final state, the scattering angles of the top and antitop quarks with respect to the jet direction are connected to their energies. This relation proves to be particularly useful in probing the charge asymmetry in the $qg$ channel, where the jet emerges from a quark at the parton level. The resulting {\it energy asymmetry} measures the charge asymmetry in terms of the difference between the top- and antitop-quark energies in the partonic CM frame. It is tailored to the LHC, where the $qg$ parton luminosity is sizeable, such that the relative $gg$ background is smaller than for the asymmetries in the $q\bar q$ channel. For the first time, we have an observable at hand that allows a measurement of the charge asymmetry in the $qg$ channel. It will be explained in detail why the energy asymmetry in the $qg$ channel is sizeable, while the previously considered forward-backward or rapidity asymmetries are tiny.

This paper is organized as follows. In Section~\ref{sec:asymmetry}, we perform a comprehensive analysis of the charge asymmetry in $t\bar t + j$ production at the parton level. We start by discussing the kinematics in the $q\bar q$ channel at an analytic level (Section~\ref{subsec:kinematics}). On this basis, we derive the incline and energy asymmetries for the $q\bar q$ and $qg$ channels (Sections~\ref{subsec:aphi} and \ref{subsec:aenergy}). The incline asymmetry is compared to the forward-backward asymmetry in Section~\ref{subsec:afb}. A summary of the new asymmetries and their kinematic features concludes the partonic analysis (Section~\ref{subsec:asparton-summary}). In Section~\ref{sec:colliders}, we study the incline and energy asymmetries at hadron colliders. At the Tevatron, the incline asymmetry is a large quantity and is expected to be accessible with the full data set (Section~\ref{subsec:iatevatron}). The energy asymmetry cannot be observed at the Tevatron. At the LHC, both the incline and the energy asymmetry can be explored (Sections~\ref{subsec:ialhc} and \ref{subsec:ealhc}). Due to the large $gg$ background, suitable cuts are indispensable to the observation of a sizeable asymmetry. We comment on the prospects to measure the new observables at the LHC with $8$ TeV data and provide predictions for the $14$ TeV run, where a high significance can be reached with a larger data set. We conclude in Section~\ref{sec:conclusions}. In Appendix~\ref{app:boosted}, we discuss the relation between the asymmetries in the $q\bar q$ and $qg$ channels. The results allow us to understand the magnitude of the respective observables in different channels.

\section{The incline and energy asymmetries}\label{sec:asymmetry}
The definition of a charge asymmetry in $t\bar t + j$ production crucially depends on the jet kinematics in the final state. Momentum conservation requires that the three-momenta of the final-state particles in the partonic process $p_1 p_2 \rightarrow t\bar t p_3$ sum to zero in the partonic CM frame, $\vec{k}_t + \vec{k}_{\bar t} + \vec{k}_{3} = 0$. The top, antitop and jet ($p_3$) momenta $\vec{k}_t$, $\vec{k}_{\bar t}$, and $\vec{k}_{3}$ thus lie in one plane, which features a certain inclination with respect to the plane spanned by the momenta of the incoming partons and the jet, $\vec{k}_{1}$, $\vec{k}_{2}$, and $\vec{k}_{3}$. Below, we will derive an asymmetry in terms of this inclination, named incline asymmetry. This new observable is particularly useful for probing the charge asymmetry in $q\bar q\rightarrow t\bar t g$ in the partonic CM frame. The incline asymmetry is essentially free from the impact of the final-state gluon direction on the angular correlation between the light quarks and the top and antitop quarks. In principle, the incline asymmetry can also be defined for the process $qg\rightarrow t\bar t q$ by boosting into a reference frame in which the initial state is antisymmetric under charge conjugation.

In addition to the incline asymmetry, the final state $t\bar t + j$ allows us to probe the charge asymmetry via an energy asymmetry, which is based on the difference between the top and antitop energies. In the process $q\bar q\rightarrow t\bar t g$, the incline asymmetry and the energy asymmetry complement each other, as they probe independent parts of the differential charge asymmetry. In $qg\rightarrow t\bar t q$, the energy asymmetry is particularly well suited to explore the charge asymmetry in the partonic CM frame. Remarkably, the energy asymmetry is the first observable that provides access to the charge asymmetry in the $qg$ channel at the LHC.

In this section, we will derive and discuss the incline asymmetry and the energy asymmetry at the parton level. We will also compare these new asymmetries with the previously investigated forward-backward asymmetry in terms of the top-quark scattering angle off the incoming quark. The jet angular distribution shows that the incline asymmetry equals the forward-backward asymmetry in its maximum, which is reached for central jet emission \cite{Berge:2012rc}. For all other scattering directions of the hard jet, the incline asymmetry is larger than the forward-backward asymmetry.

\subsection{Differential charge asymmetry}\label{subsec:kinematics}
\begin{figure}[!t]
\begin{center}
\includegraphics[scale=0.4]{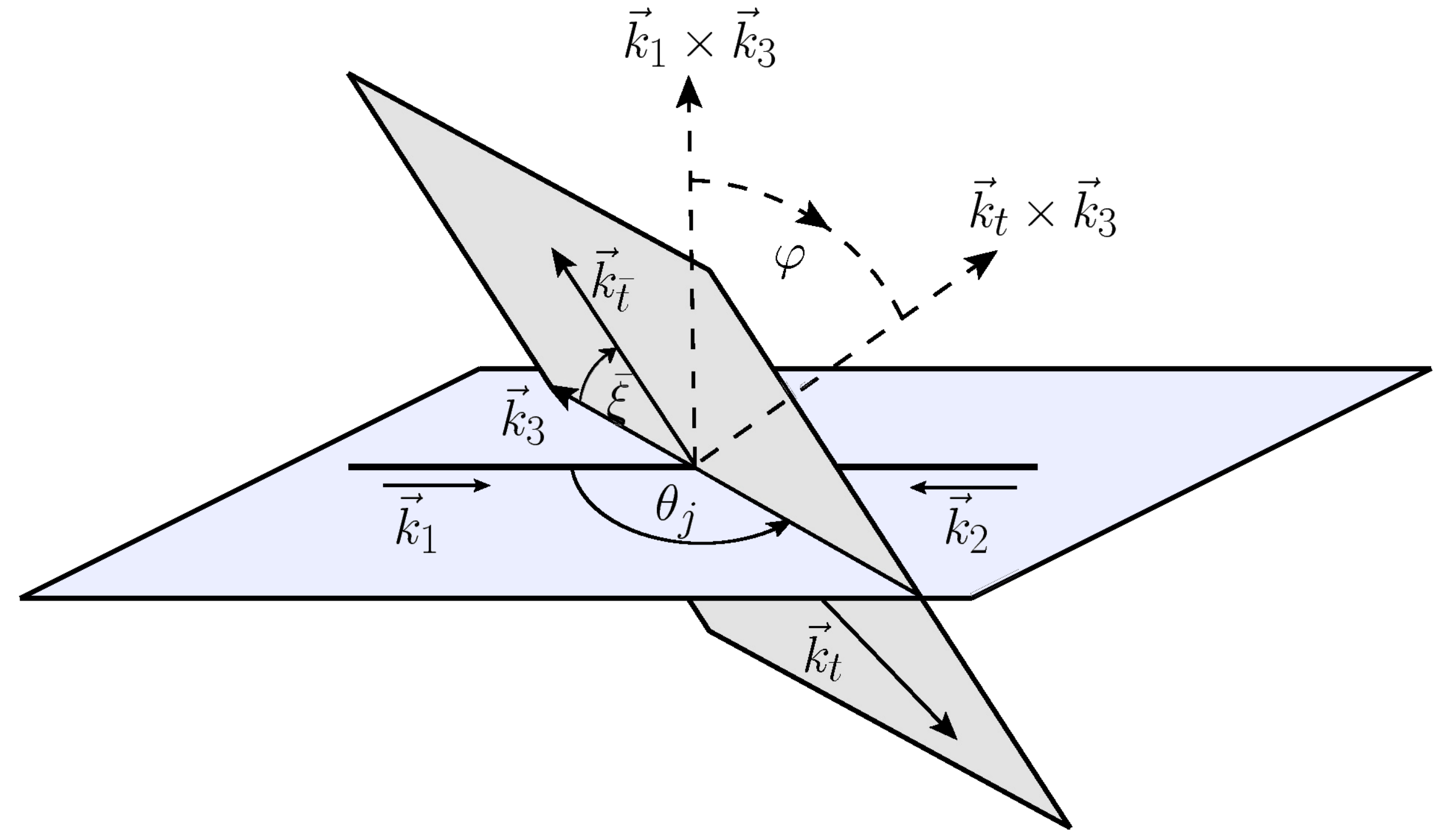}
\end{center}
\vspace*{-1cm}
\begin{center} 
  \parbox{15.5cm}{\caption{\label{fig:kinematics}Kinematics of the process $p_1 p_2\rightarrow t\bar t p_3$. The definition of the inclination angle $\varphi$ is given in (\ref{eq:angles}).}}
\end{center}
\end{figure}

To describe the parton kinematics in a way that captures the relevant features of the incline and energy asymmetries, we use the parameterization from \cite{Hahn:1998yk}. In this parameterization, the three-body final state in the process $p_1 p_2 \rightarrow t \bar t p_3$ is described by five variables in the partonic CM system: the top-quark and jet\footnote{We use the term ``jet'' at the parton level to refer to a light quark or a gluon.} energies $E_t$ and $E_j$, the jet scattering angle $\theta_j$ off the direction of the incoming parton $p_1$, and the inclination angle $\varphi$ between the planes spanned by the three-momenta $(\vec{k}_{1},\vec{k}_{3})$ and $(\vec{k}_t,\vec{k}_{3})$. The fifth degree of freedom, an azimuthal rotation of the final state around the beam axis, is irrelevant for our purposes. The kinematical setup is visualized in Figure~\ref{fig:kinematics}.
The four-momenta of the partons are given by
\begin{eqnarray}\label{eq:parametrization}
k_1 & = & \sqrt{s}/2\,(1,0,0,1)\,,\quad k_2 = \sqrt{s}/2\,(1,0,0,-1)\,,\quad k_3 = E_j\,(1,\sin\theta_j,0,\cos\theta_j)\,,\nonumber\\
k_t & = & (E_t,\sqrt{E_t^2-m_t^2}\cdot\vec{e}_t)\,,\quad k_{\bar t} = k_1 + k_2 - k_3 - k_t\,,\quad \text{with}\nonumber\\
\vec{e}_t & = & \left(\begin{array}{c}
\sin\theta_{t} \cos\phi_{t}\\
\sin\theta_{t} \sin\phi_{t}\\
\cos\theta_{t}\end{array}\right)\,\,=\,\,\left(\begin{array}{ccc}
\cos\theta_j & 0 & \sin\theta_j\\
0 & 1 & 0\\
-\sin\theta_j & 0 & \cos\theta_j\end{array}\right)\cdot\left(\begin{array}{c}
-\cos\varphi\sin\xi\\
-\sin\varphi\sin\xi\\
\cos\xi\end{array}\right)\label{eq:Normalized_Top_Vec_def}\\
 & = & \left(\begin{array}{c}
-\cos\theta_j\cos\varphi\sin\xi+\sin\theta_j\cos\xi\\
-\sin\varphi\sin\xi\\
\sin\theta_j\cos\varphi\sin\xi+\cos\theta_j\cos\xi\end{array}\right)\,, \nonumber 
\end{eqnarray} 
where $\sqrt{s}$ is the partonic CM energy and $m_t$ is the top-quark mass. The jet scattering angle $\theta_j$ and the inclination angle $\varphi$ fix the position of the final-state plane $(t,\bar t,p_3)$ with respect to the plane $(p_1,p_2,p_3)$. They are defined by\footnote{Note that $\varphi$ corresponds to $\eta + \pi$ in \cite{Hahn:1998yk}.}
\begin{eqnarray}\label{eq:angles}
\cos\theta_j & = & \vec{k}_1\cdot \vec{k}_3/(|\vec{k}_1||\vec{k}_3|)\,,\quad\ \, \theta_j \in [0,\pi]\,,\\\nonumber
\cos\varphi & = & \vec{n}_{13}\cdot\vec{n}_{t3}\,,\qquad\qquad\quad \varphi \in [0,2\pi]\,, 
\end{eqnarray}
with the normal vectors of the planes $(p_1,p_2,p_3)$ and $(t,\bar t,p_3)$, $\vec{n}_{13} = (\vec{k}_1\times\vec{k}_3)/|\vec{k}_1\times\vec{k}_3|$ and $\vec{n}_{t3} = (\vec{k}_t\times\vec{k}_3)/|\vec{k}_t\times\vec{k}_3|$. The angle $\xi$ between the top-quark and jet momenta is fixed in terms of $E_t$, $E_j$ and $\sqrt{s}$ by the relation
\begin{eqnarray}\label{eq:cosxi}
\cos\xi & = & \frac{\vec{k}_t\cdot \vec{k}_3}{|\vec{k}_t||\vec{k}_3|} = \frac{s-2\sqrt{s}\,(E_t+E_j)+2E_tE_j}{2E_j\sqrt{E_t^{2}-m_t^{2}}}\,,\qquad \xi \in [0,\pi]\,. 
\end{eqnarray}
The range of the energies $E_t$ and $E_j$ is independent from $\theta_j$ and $\varphi$ and given in~\cite{Hahn:1998yk}. The angle $\bar{\xi}$ between the antitop-quark and jet momenta is obtained by changing $E_t$ to the antitop-quark energy $E_{\bar t} = \sqrt{s} - E_t - E_j$ in (\ref{eq:cosxi}),
\begin{eqnarray}\label{eq:cosxibar}
\cos\bar{\xi} & = & \frac{\vec{k}_{\bar t}\cdot \vec{k}_3}{|\vec{k}_{\bar t}||\vec{k}_3|} = \frac{s-2\sqrt{s}\,(E_{\bar t}+E_j)+2E_{\bar t}E_j}{2E_j\sqrt{E_{\bar t}^{2}-m_t^{2}}}\,,\qquad \bar{\xi} \in [0,\pi]\,. 
\end{eqnarray}

For $t\bar t + j$ production at hadron colliders, the three partonic processes in QCD at LO are the quark-antiquark channel $q\bar q\rightarrow t\bar t g$, where $\{p_1,p_2,p_3\} = \{q,\bar q,g\}$, the quark-gluon channel $qg\rightarrow t\bar t q$, where $\{p_1,p_2,p_3\} = \{q,g,q\}$, and the gluon-gluon channel $gg\rightarrow t\bar t g$. The $gg$ channel does not contribute to the charge asymmetry, but is an important background at the LHC. The parameterization introduced above allows us to control the infrared and collinear behavior of the partonic cross sections in the limits $E_j\rightarrow 0$ and $\theta_j\rightarrow 0,\pi$. For later purposes, it is important to bear in mind that the charge asymmetry in the $q\bar q$ channel has an infrared divergence, whereas the $qg$ contribution is finite for $E_j\rightarrow 0$. The symmetric cross section for $t\bar t + j$ production, which normalizes the asymmetry, exhibits both soft and collinear divergences. The size of the observable charge asymmetry thus depends significantly on $E_j$ and $\theta_j$, and on the cuts being introduced to regularize the divergences.

The differential charge-symmetric cross section $\text{d}\hat{\sigma}_S$ and the charge-asymmetric cross section $\text{d}\hat{\sigma}_A$ at the parton level are defined as
\begin{eqnarray}\label{eq:sigmaadef}
\text{d}\hat{\sigma}_S & = & \text{d}\hat{\sigma}(p_1 p_2\rightarrow t(k_t)\,\bar t(k_{\bar t})\,p_3) + \text{d}\hat{\sigma}(p_1 p_2\rightarrow \bar t(k_t)\,t(k_{\bar t})\,p_3)\,,\\
\text{d}\hat{\sigma}_A & = & \text{d}\hat{\sigma}(p_1 p_2\rightarrow t(k_t)\,\bar t(k_{\bar t})\,p_3) - \text{d}\hat{\sigma}(p_1 p_2\rightarrow \bar t(k_t)\,t(k_{\bar t})\,p_3)\,.\nonumber
\end{eqnarray}
Since the $q\bar q$ initial state is antisymmetric under charge conjugation and the jet angular {di\-stri\-bu\-tion} is symmetric, the analytic expression of the charge asymmetries is most transparent in this channel. We will thus primarily focus on the process $q\bar q\rightarrow t\bar t j$, when we introduce the new asymmetries at the parton level. In the $qg$ channel, the initial state is neither symmetric nor antisymmetric under charge conjugation. Furthermore, the jet distribution in $qg\rightarrow t\bar t j$ is not forward-backward symmetric, but the jet is preferentially emitted in the direction of the incoming quark. As we aim at examining the features of the charge asymmetry in terms of top and antitop energies, we substitute the jet energy $E_j$ with the antitop energy $E_{\bar t}$ in our parameterization in (\ref{eq:parametrization}). In terms of the inclination angle $\varphi$, the jet scattering angle $\theta_j$, and the top and antitop energies $E_t$ and $E_{\bar t}$, the differential charge asymmetry in $q\bar q\rightarrow t\bar t j$ exhibits the following structure,
\begin{eqnarray}\label{eq:sigmaa}
\frac{\text{d}\hat{\sigma}_A(q\bar q\rightarrow t\bar t j)}{\text{d}\varphi\,\text{d}\theta_j\,\text{d}E_t\,\text{d}E_{\bar t}} & = & \frac{\alpha_s^3}{4\pi s}\frac{d_{abc}^2}{16N_C^2}\Big\{-[N_1 + \sin^2\theta_j\,(N_1^j + \cos^2\varphi\,N_1^{\varphi})]\,\cos\varphi\\
 & & \hspace{4cm} {}+\,[N_2 + \cos^2\varphi\,N_2^{\varphi}]\,\sin\theta_j\cos\theta_j\Big\}\,,\nonumber 
\end{eqnarray}
with the $SU(3)$ color factors $N_C=3$ and $d_{abc}^2=40/3$, and the coefficients $\{N_i\} = \{N_i(E_t,E_{\bar t})\}$ given by
\begin{eqnarray}\label{eq:coefficients}
N_1 & = & \frac{16}{D}\,\frac{C}{(\sqrt{s}-E_t-E_{\bar t})^3}\,\Big\{4(E_t + E_{\bar t})^2[3s/2 - 2m_t^2 + E_t^2 + E_{\bar t}^2]\\
   & & {}- 4\sqrt{s}(E_t+E_{\bar t})[s - m_t^2 + E_t^2 + E_{\bar t}^2 + (E_t+E_{\bar t})^2(1-m_t^2/s)] + s[s + 2(E_t^2+E_{\bar t}^2)]\Big\}\,,\nonumber\\
N_1^j & = & \frac{16}{D}\,\frac{C}{(\sqrt{s}-E_t-E_{\bar t})^3}\,\Big\{-2(E_t + E_{\bar t})^2[3s - 6m_t^2 + E_t^2 + E_{\bar t}^2]\nonumber\\
   & & {}+ 4\sqrt{s}(E_t+E_{\bar t})[s - 3m_t^2 + (E_t+E_{\bar t})^2(1-m_t^2/s)] - s[s - 4m_t^2]\Big\}\,,\nonumber\\
N_1^{\varphi} & = & \frac{16}{D}\,\frac{C^{3}}{(\sqrt{s}-E_t-E_{\bar t})^3}\,,\nonumber\\
N_2 & = & \frac{64}{D}\,m_t^2\,(E_t - E_{\bar t})\,,\qquad N_2^{\varphi} \,=\, \frac{32}{D}\,\frac{C^2}{(\sqrt{s}-E_t-E_{\bar t})^3}\,(E_t^2 - E_{\bar t}^2)\,,\nonumber 
\end{eqnarray}
where
\begin{eqnarray}
D & = & s\left[2(E_t + E_{\bar t}) - \sqrt{s}\right](\sqrt{s} - 2E_t)(\sqrt{s}-2E_{\bar t})\,,\nonumber\\
C & = & 2\,\sqrt{-(E_t+E_{\bar t})^2(s+m_t^2) + \sqrt{s}(E_t + E_{\bar t})[s + 2m_t^2 + 2E_tE_{\bar t}] - s[s/4+m_t^2+E_tE_{\bar t}]}\,.\nonumber 
\end{eqnarray}
The first term of the distribution $\text{d}\hat{\sigma}_A(q\bar q\rightarrow t\bar t j)$ in (\ref{eq:sigmaa}) is antisymmetric in $\cos\varphi$, whereas the second term is antisymmetric in $\cos\theta_j$. An asymmetry in terms of $\cos\varphi$, the incline asymmetry, thus probes the set of coefficients $\{N_1\}=\{N_1,N_1^{j},N_1^{\varphi}\}$. While $\{N_1\}$ are symmetric under $E_t\leftrightarrow E_{\bar t}$, the coefficients $\{N_2\}=\{N_2,N_2^{\varphi}\}$ are antisymmetric under $E_t\leftrightarrow E_{\bar t}$. An asymmetry in terms of the energy difference $\Delta E = E_t - E_{\bar t}$, the energy asymmetry, thus probes the complementary part of the differential asymmetry via $\{N_2\}$. However, in the $q\bar q$ channel, the energy asymmetry vanishes when integrated over $\theta_j$. One has to construct a two-fold asymmetry in terms of $\Delta E$ and $\theta_j$ to obtain a non-vanishing observable. In the $q\bar q$ channel, the asymmetries in $\cos\varphi$ and $\Delta E$ provide complementary information about the charge-asymmetric cross section $\text{d}\hat{\sigma}_A(q\bar q\rightarrow t\bar tj)$ in (\ref{eq:sigmaa}). In the $qg$ channel, this complementarity is lifted, because the jet originates from a quark rather than from a gluon.

\subsection{Incline asymmetry at the parton level}\label{subsec:aphi}
In the $q\bar q$ channel, the incline asymmetry probes the charge asymmetry $\text{d}\hat{\sigma}_A(q\bar q\rightarrow t\bar tj)$ via the inclination angle $\varphi$ between the planes $(t,\bar t,j)$ and $(q,\bar q,j)$. We define the differential partonic incline asymmetry in terms of the jet scattering angle as
\begin{eqnarray}\label{eq:sigmaa-phi}
\frac{\text{d}\hat{\sigma}_A^{\varphi}}{\text{d}\theta_j} & \equiv & \frac{\text{d}\hat{\sigma}(\cos\varphi > 0)}{\text{d}\theta_j} - \frac{\text{d}\hat{\sigma}(\cos\varphi < 0)}{\text{d}\theta_j}\,. 
\end{eqnarray}
The integrated partonic incline asymmetry is then given by 
\begin{eqnarray}\label{eq:aphi}
\hat{A}^{\varphi} & \equiv & \frac{\hat{\sigma}_A^{\varphi}}{\hat{\sigma}_S} \,=\, \frac{\int_{0}^{\pi} \text{d}\theta_j\,\text{d}\hat{\sigma}_A^{\varphi}}{\hat{\sigma}_S} \,={} -\frac{1}{\hat{\sigma}_S}\frac{\alpha_s^3}{4\pi s}\frac{d_{abc}^2}{16N_C^2}\Big[2\pi\,\widetilde{N}_1 + \pi\, \widetilde{N}_1^j + \frac{2\pi}{3}\,\widetilde{N}_1^{\varphi}\Big]\,, 
\end{eqnarray}
where $\hat{\sigma}_S$ is the total partonic cross section. Here and in the following sections, the jet distribution $\text{d}\hat{\sigma}_A/\text{d}\theta_j$ is denoted by $\text{d}\hat{\sigma}_A$ under the integrand, and the variables $\varphi$, $E_{t}$, and $E_{\bar t}$ are implicitly integrated over. The coefficients $\widetilde{N}_1$, $\widetilde{N}_1^j$ and $\widetilde{N}_1^{\varphi}$ are obtained from (\ref{eq:coefficients}) by integrating $\{N_1\}$ over $E_t$ and $E_{\bar t}$.  The prefactors are due to the integration over $\varphi$ and $\theta_j$. As we observe in (\ref{eq:aphi}), the incline asymmetry $\hat{A}^{\varphi}$ is sensitive to a combination of $\{\widetilde{N}_1\}$, namely to that part of the differential charge asymmetry which is symmetric in the top-antitop energy difference $\Delta E$.

\begin{figure}[!t]
\begin{center}
\includegraphics[height=5.5cm]{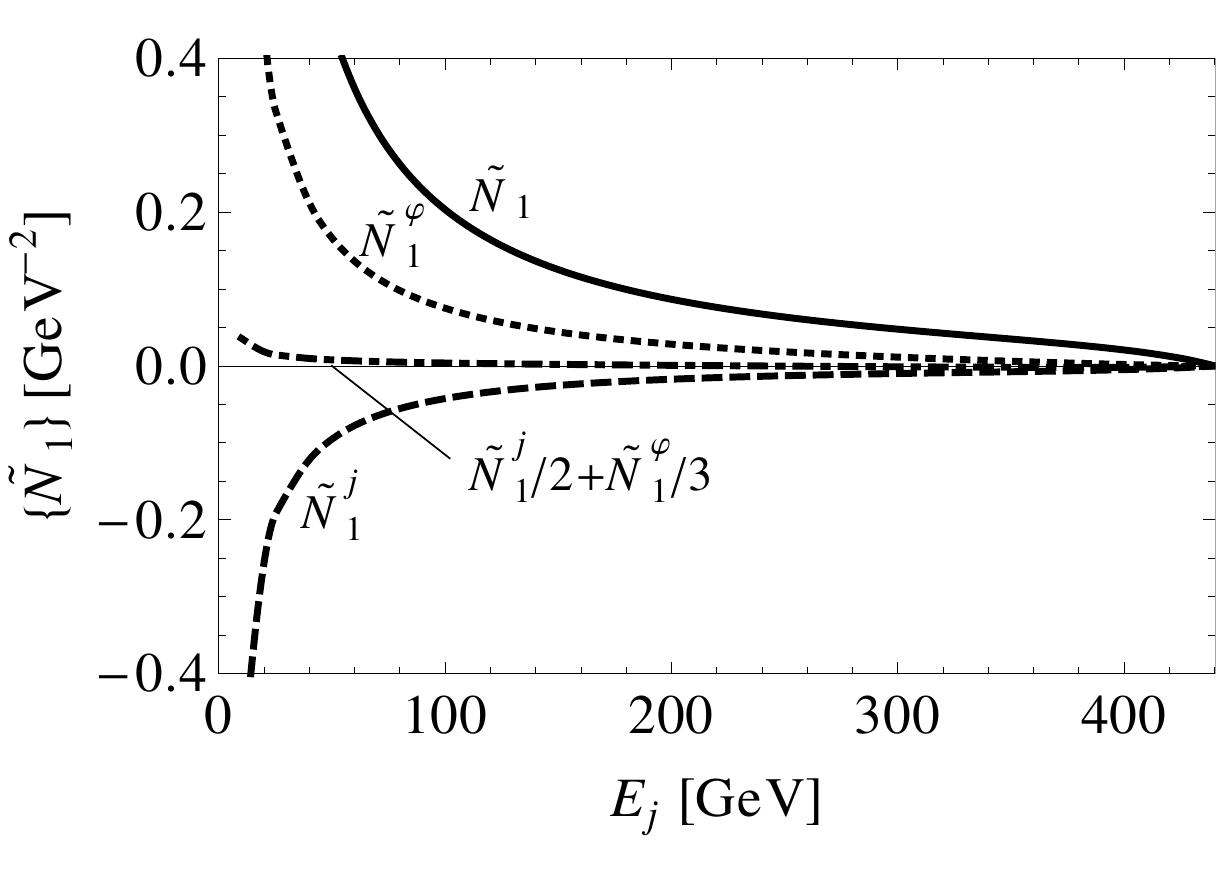}
\hspace*{0.5cm}
\includegraphics[height=5.5cm]{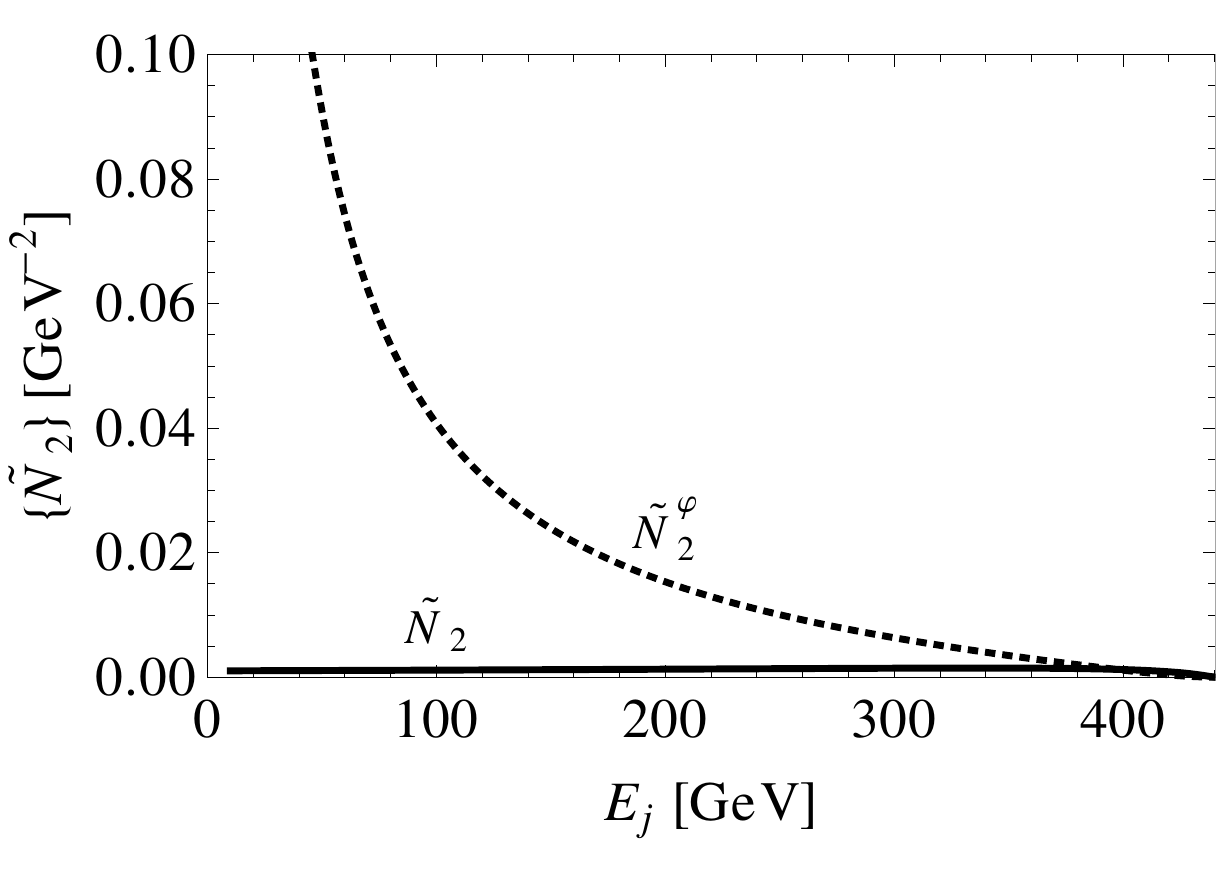}
\end{center}
\vspace*{-1cm}
\begin{center} 
  \parbox{15.5cm}{\caption{\label{fig:Ni} Contributions to the incline asymmetry $\hat{\sigma}_A^{\varphi}$ (left) and the energy asymmetry $\hat{\sigma}_A^{E,j}$ (right) for the $q\bar q$ channel in terms of the coefficients $\{\widetilde{N}_1\}$ and $\{\widetilde{N}_2\}$, as functions of the jet energy $E_j$ for $\sqrt{s} = 1\,\text{TeV}$.}} 
\end{center}
\end{figure}

These different contributions to $\hat{A}^{\varphi}$ are displayed in the left panel of Figure~\ref{fig:Ni} as functions of the jet energy $E_j$ for a partonic CM energy of $\sqrt{s} = 1\,\text{TeV}$.\footnote{In $\{\widetilde{N}_1(E_t,E_{\bar t})\}$, the dependence on $E_{\bar t}$ has been substituted with $E_j = \sqrt{s} - E_t - E_{\bar t}$, and the remaining dependence on $E_t$ has been integrated out.}
We observe that all coefficients $\widetilde{N}_1$ diverge in the infrared limit $E_j\rightarrow 0$. The main contribution to the asymmetry $\hat{A}^{\varphi}$ thus stems from soft gluon emission. Numerically, the coefficient $\widetilde{N}_1$ is larger than $\widetilde{N}_1^j$ and $\widetilde{N}_1^{\varphi}$, which partially cancel in the combination $\widetilde{N}_1^j/2 + \widetilde{N}_1^{\varphi}/3$ that enters $\hat{A}^{\varphi}$. From (\ref{eq:sigmaa}), it is apparent that the contribution of $\widetilde{N}_1$ to the asymmetry is independent from $\theta_j$. Due to this behavior, the incline asymmetry is insensitive to the jet scattering angle. The incline asymmetry therefore provides optimal access to the differential charge asymmetry $\text{d}\hat{\sigma}_A(q\bar q\rightarrow t\bar t j)/\text{d}\varphi\,\text{d}\theta_j\,\text{d}E_t\,\text{d}E_{\bar t}$, independently of the jet direction.

In the $qg$ channel, the incline asymmetry as defined in (\ref{eq:sigmaa-phi}) and (\ref{eq:aphi}) is based on the inclination between the planes $(t,\bar t,q)$ and $(q,g,q)$. The inclination angle $\varphi$ is now defined with respect to the $quark$-jet momentum and therefore inappropriate to measure the angular correlation between the quarks in the initial and final states. A definition of the incline asymmetry with respect to the gluon momentum $\vec{k}_2$ is not possible in the partonic CM frame, because in general $\vec{k}_t$, $\vec{k}_{\bar t}$ and $\vec{k}_2$ do not lie in one plane. By boosting into the $t\bar t$ rest frame, the incline asymmetry in the $qg$ channel can be defined on the basis of a charge-antisymmetric $t\bar t$ initial state, in analogy to the $q\bar q$ channel. We will only briefly comment on this possibility in Appendix~\ref{app:boosted}, since the energy asymmetry will prove more useful as an observable of the charge asymmetry in the $qg$ channel.

\subsection{Energy asymmetry at the parton level}\label{subsec:aenergy}
The energy asymmetry is based on the difference between the top and antitop quarks' energies, $\Delta E = E_t - E_{\bar t}$. We define the differential energy asymmetry as a function of the jet angle by
\begin{eqnarray}\label{eq:sigmaa-energy}
\frac{\text{d}\hat{\sigma}_A^E}{\text{d}\theta_j} & \equiv & \frac{\text{d}\hat{\sigma}(\Delta E > 0)}{\text{d}\theta_j} - \frac{\text{d}\hat{\sigma}(\Delta E < 0)}{\text{d}\theta_j}\,. 
\end{eqnarray}
The energy difference $\Delta E$ is connected to the kinematics in the final-state plane $(t,\bar t,j)$ by energy and momentum conservation. In particular, the angles $\xi$ and $\bar{\xi}$ of the top and antitop quarks with respect to the jet momentum, defined in (\ref{eq:cosxi}) and (\ref{eq:cosxibar}), are fixed by the energies $E_t$ and $E_{\bar t} = \sqrt{s} - E_t - E_j$. For $\Delta E > 0$, one has $\cos\bar{\xi} > \cos\xi$, and for $\Delta E < 0$, one has $\cos\bar{\xi} < \cos\xi$. The energy asymmetry $\text{d}\hat{\sigma}_A^E/\text{d}\theta_j$ can therefore be interpreted as an asymmetry of the top and antitop scattering angles with respect to the jet direction,
\begin{eqnarray}\label{eq:energy-angular}
\frac{\text{d}\hat{\sigma}_A^E}{\text{d}\theta_j} & = & {}-\frac{\text{d}\hat{\sigma}_A^{\xi}}{\text{d}\theta_j} \,\equiv\, {}-\Big(\frac{\text{d}\hat{\sigma}(\cos\xi > \cos\bar{\xi})}{\text{d}\theta_j} - \frac{\text{d}\hat{\sigma}(\cos\xi < \cos\bar{\xi})}{\text{d}\theta_j}\Big)\,,\qquad \xi,\bar{\xi}\in [0,\pi]\,. 
\end{eqnarray}
In Appendix~\ref{app:boosted}, we exploit this correspondence and relate the energy asymmetry in the $q\bar q$ channel to the top-angle asymmetry in the $qg$ channel and vice versa. This relation allows us to understand the magnitude of the asymmetries in the respective channels.

For the $q\bar q$ channel, the energy-asymmetric part of $\text{d}\hat{\sigma}_A(q\bar q\rightarrow t\bar t j)$ is encoded in the coefficients $\{N_2\}$ given in (\ref{eq:sigmaa}) and (\ref{eq:coefficients}). It accompanies an asymmetry in the (gluon-)jet scattering angle $\theta_j$. The energy asymmetry $\text{d}\hat{\sigma}_A^E/\text{d}\theta_j$ is thus antisymmetric under $\theta_j \leftrightarrow \pi - \theta_j$ and {va\-nishes} when integrated over $\theta_j \in [0,\pi]$. To construct a non-vanishing observable, one needs to consider a two-fold asymmetry in terms of $\Delta E$ and $\theta_j$. We thus define the partonic energy asymmetry for the process $q\bar q\rightarrow t\bar tj$ as
\begin{eqnarray}\label{eq:energy-asymmetry-qqb}
\hat{A}^{E,j} & \equiv & \frac{\hat{\sigma}_A^{E,j}}{\hat{\sigma}_S} \,=\, \frac{\int_{0}^{\pi/2} \text{d}\theta_j\,\text{d}\hat{\sigma}_A^{E} - \int_{\pi/2}^{\pi} \text{d}\theta_j\,\text{d}\hat{\sigma}_A^E}{\hat{\sigma}_S} \,=\, \frac{1}{\hat{\sigma}_S}\frac{\alpha_s^3}{4\pi s}\frac{d_{abc}^2}{16N_C^2}\Big[2\pi\widetilde{N}_2 + \pi\,\widetilde{N}_2^{\varphi}\Big]\,. 
\end{eqnarray}
The coefficients $\{\widetilde{N}_2\}$ are obtained from (\ref{eq:coefficients}) by integrating $\{N_2\}$ over $E_t$ and $E_{\bar t} < E_t$. The factors $2\pi$ and $\pi$ in front of $\widetilde{N}_2 $ and $\widetilde{N}_2^{\varphi}$, respectively, stem from the integration 
over $\varphi$. By comparing the coefficients $\{\widetilde{N}_i\}$ in (\ref{eq:aphi}) and (\ref{eq:energy-asymmetry-qqb}), it becomes once more apparent that the energy asymmetry $\hat{A}^{E,j}$ is independent from and complementary to the incline asymmetry $\hat{A}^{\varphi}$.

\begin{figure}[!t]
\begin{center}
\includegraphics[height=5.5cm]{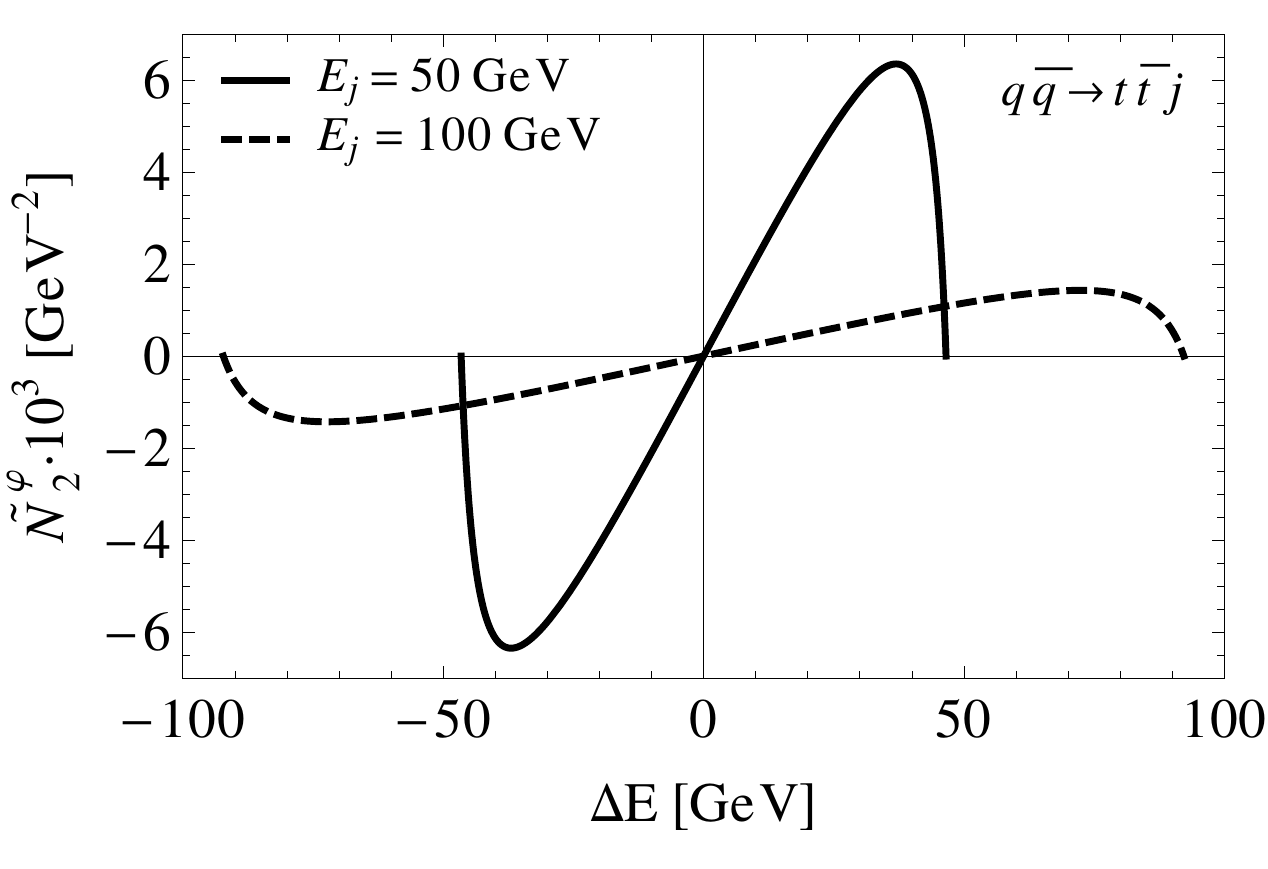}
\hspace*{0.5cm}
\includegraphics[height=5.5cm]{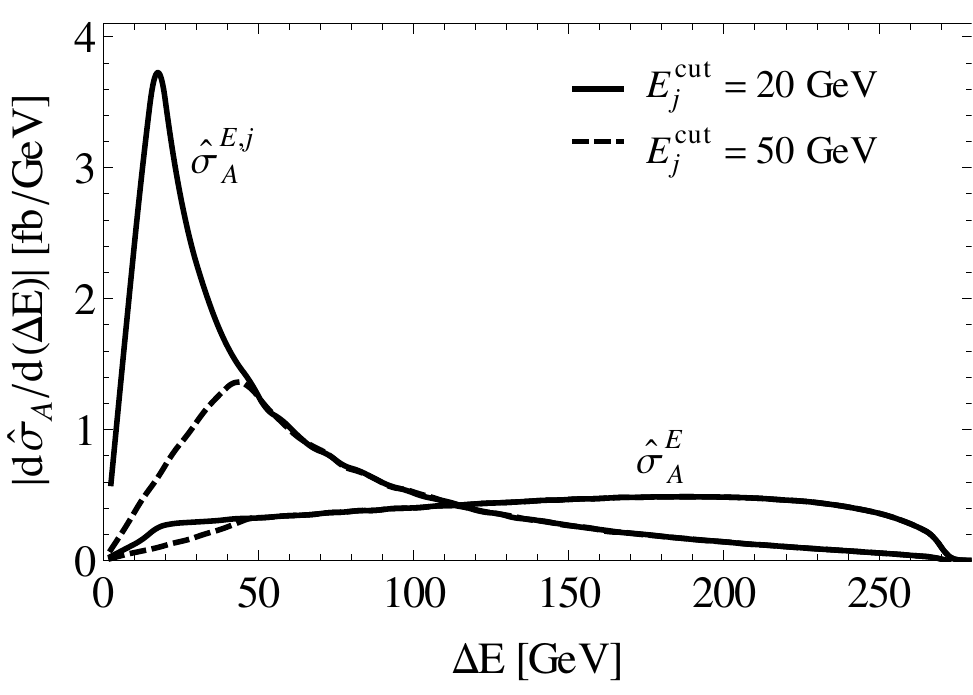}
\end{center}
\vspace*{-1cm}
\begin{center} 
  \parbox{15.5cm}{\caption{\label{fig:N2phiDE} Contributions to the partonic energy asymmetries $\hat{\sigma}_A^{E,j}$ ($q\bar q$ channel) and $\hat{\sigma}_A^{E}$ ($qg$ channel) as functions of $\Delta E$ at $\sqrt{s} = 1\,\text{TeV}$. Left: dominant coefficient $\widetilde{N}_2^{\varphi}$ ($q\bar q$) for fixed values of the jet energy $E_j$. Right: absolute differential asymmetry $|\text{d}\hat{\sigma}_A^{E,(j)}/\text{d}(\Delta E)|$ for $q\bar q$ and $qg$ with jet energy cuts $E_j^{\text{cut}} = 20,50\,\text{GeV}$.}}  
\end{center}
\end{figure}

In order to examine the kinematic features of $\hat{A}^{E,j}$ in the $q\bar{q}$ channel, we study the coefficients $\{N_2(E_t,E_{\bar t})\}$ as functions of $E_j \sim E_t + E_{\bar t}$ and $\Delta E = E_t - E_{\bar t}$. The dependence of $\widetilde{N}_2$ and $\widetilde{N}_2^{\varphi}$ on the jet energy $E_j$ is shown in the right panel of Figure~\ref{fig:Ni}. Since $\widetilde{N}_2$ is finite for $E_j\to 0$, the energy asymmetry is governed by $\widetilde{N}_2^{\varphi}$, unless $E_j$ is close to its maximum. By comparing the scales of the coefficients $\{\widetilde{N}_1\}$ and $\{\widetilde{N}_2\}$ in Figure~\ref{fig:Ni}, the energy asymmetry $\hat{A}^{E,j}$ is expected to be much smaller in magnitude than the incline asymmetry $\hat{A}^{\varphi}$. The dependence of the dominant coefficient $\widetilde{N}_2^{\varphi}$ on $\Delta E$ is displayed in Figure~\ref{fig:N2phiDE}, left, for fixed values of the jet energy, $E_j = 50\,\text{GeV}$ (plain) and $E_j = 100\,\text{GeV}$ (dashed). The range of $\Delta E$ is limited for small $E_j$ due to momentum conservation in the final state. Within this range, the maximum of $\widetilde{N}_2^{\varphi}$ is reached for large $\Delta E$, corresponding to $\Delta E \approx E_j$ for not too large $E_j$.

In the $qg$ channel, the jet distribution $\text{d}\hat{\sigma}_A^E/\text{d}\theta_j$ in (\ref{eq:sigmaa-energy}) is not antisymmetric under $\theta_j\leftrightarrow \pi - \nolinebreak[4]\theta_j$, because the initial state is not charge-antisymmetric as in the $q\bar q$ channel. The differential energy asymmetry in (\ref{eq:sigmaa-energy}) can therefore simply be integrated over the jet scattering angle $\theta_j$. The partonic energy asymmetry for $qg\rightarrow t\bar t j$ is then given by
\begin{eqnarray}\label{eq:energy-asymmetry-qg}
 \hat{A}^E & \equiv & \frac{\hat{\sigma}_A^E}{\hat{\sigma}_S} = \frac{\int_0^{\pi} \text{d}\theta_j\,\text{d}\hat{\sigma}_A^E}{\hat{\sigma}_S}\,. 
\end{eqnarray}
Importantly, neither the jet direction nor the initial-state quark direction enter the definition of the charge asymmetry $\hat{A}^E$. This feature will be of great use to construct an observable at the LHC.

The dependence of the energy asymmetry on $\Delta E$ is different for the $q\bar q$ and $qg$ channels. In Figure~\ref{fig:N2phiDE}, right, we show the absolute differential asymmetries $|\text{d}\hat{\sigma}_A^{E,j}/\text{d}(\Delta E)|$ and $|\text{d}\hat{\sigma}_A^{E}/\text{d}(\Delta E)|$ for lower cuts on the jet energy, $E_j^{\text{cut}} = 20\,\text{GeV}$ (plain) and $E_j^{\text{cut}} = 50\,\text{GeV}$ (dashed). The dependence on the jet angle $\theta_j$ has been integrated out. Since the asymmetry in the $q\bar q$ channel has an infrared divergence for $E_j\rightarrow 0$, the main contribution to the asymmetry stems from small jet energies close to the cut. The range of $\Delta E$ is therefore limited to small values $\Delta E \approx E_j^{\text{cut}}$ (see Figure~\ref{fig:N2phiDE}, left), such that $\text{d}\hat{\sigma}_A^{E,j}/\text{d}(\Delta E)$ depends strongly on the jet energy cut. On the contrary, in the $qg$ channel, which is finite for $E_j\rightarrow 0$, the distribution $\text{d}\hat{\sigma}_A^{E}/\text{d}(\Delta E)$ is rather flat. A jet energy cut has only a minor effect on the integrated asymmetry $\hat{\sigma}_A^E$. These kinematic features will be important for the definition of cuts for the hadronic observables.

\subsection{Comparison with the forward-backward asymmetry}\label{subsec:afb}
The incline asymmetry $\hat{A}^{\varphi}$ in the $q\bar q$ channel is kinematically connected to the asymmetry $\hat{A}^{\theta_t}$ in terms of the top-quark scattering angle $\theta_t$. The latter has been measured as a forward-backward asymmetry in inclusive $t\bar t$ production at the Tevatron and extensively studied in the literature. We compare both asymmetries in $t\bar t + j$ production and demonstrate that $\hat{A}^{\varphi}$ is superior to $\hat{A}^{\theta_t}$ in its sensitivity to the differential cross section $\text{d}\hat{\sigma}_A(q\bar q\rightarrow t\bar t j)$. 
The partonic top-angle asymmetry is defined by\footnote{This definition applies to $q\bar q\rightarrow t\bar t g$, as well as to $qg\rightarrow t\bar t q$.}
\begin{eqnarray}\label{eq:top-angle-as}
\hat{A}^{\theta_t} & \equiv & \frac{\hat{\sigma}_A^{\theta_t}}{\hat{\sigma}_S} \,=\, \frac{\int_0^{\pi}\text{d}\theta_j\,\text{d}\hat{\sigma}_A^{\theta_t}}{\hat{\sigma}_S}\,,
\end{eqnarray}
with the jet distribution
\begin{eqnarray}\label{eq:top-angle-distribution}
\frac{\text{d}\hat{\sigma}_A^{\theta_t}}{\text{d}\theta_j} & \equiv & \frac{\text{d}\hat{\sigma}(\cos\theta_t > \cos\theta_{\bar t})}{\text{d}\theta_j} - \frac{\text{d}\hat{\sigma}(\cos\theta_t < \cos\theta_{\bar t})}{\text{d}\theta_j}\,.
\end{eqnarray}
In inclusive $t\bar t$ production, the jet direction is not detected. In this case, the top-angle asymmetry $\hat{\sigma}_A^{\theta_t}$ in the $q\bar q$ channel is equal to the forward-backward asymmetry $\hat{\sigma}_{\text{FB}} = \hat{\sigma}(\cos\theta_t >\nolinebreak[4] 0) - \hat{\sigma}(\cos\theta_t < 0)$. In $t\bar t + j$ production, this relation holds true as long as the jet direction is integrated over, because the jet distribution in $q\bar q\rightarrow t\bar t j$ is symmetric. The differential distributions $\text{d}\hat{\sigma}_A^{\theta_t}/\text{d}\theta_j$ and $\text{d}\hat{\sigma}_{\text{FB}}/\text{d}\theta_j$ are equal only for $\theta_j = \pi/2$. For fixed angles $\theta_j\neq \pi/2$, the forward-backward asymmetry $\hat{\sigma}_{\text{FB}}$ of the top quark is to a significant extent due to momentum conservation in the final state, which blurs the charge asymmetry.\footnote{In the $qg$ channel, there is no relation between $\hat{\sigma}_A^{\theta_t}$ and $\hat{\sigma}_{\text{FB}}$.  The top-quark forward-backward asymmetry originates in large part from the non-symmetric jet distribution.} In $t\bar t + j$ production, we will thus refer to the top-angle asymmetry as defined in (\ref{eq:top-angle-as}) and (\ref{eq:top-angle-distribution}). This definition is free from kinematical asymmetries caused by the jet direction. At the hadron level, the top-angle asymmetry is equal to the rapidity asymmetry defined later in (\ref{eq:afbt}), which has been measured in inclusive $t\bar t$ production by the Tevatron experiments.

\begin{figure}[!t]
\begin{center}
\includegraphics[height=5.5cm]{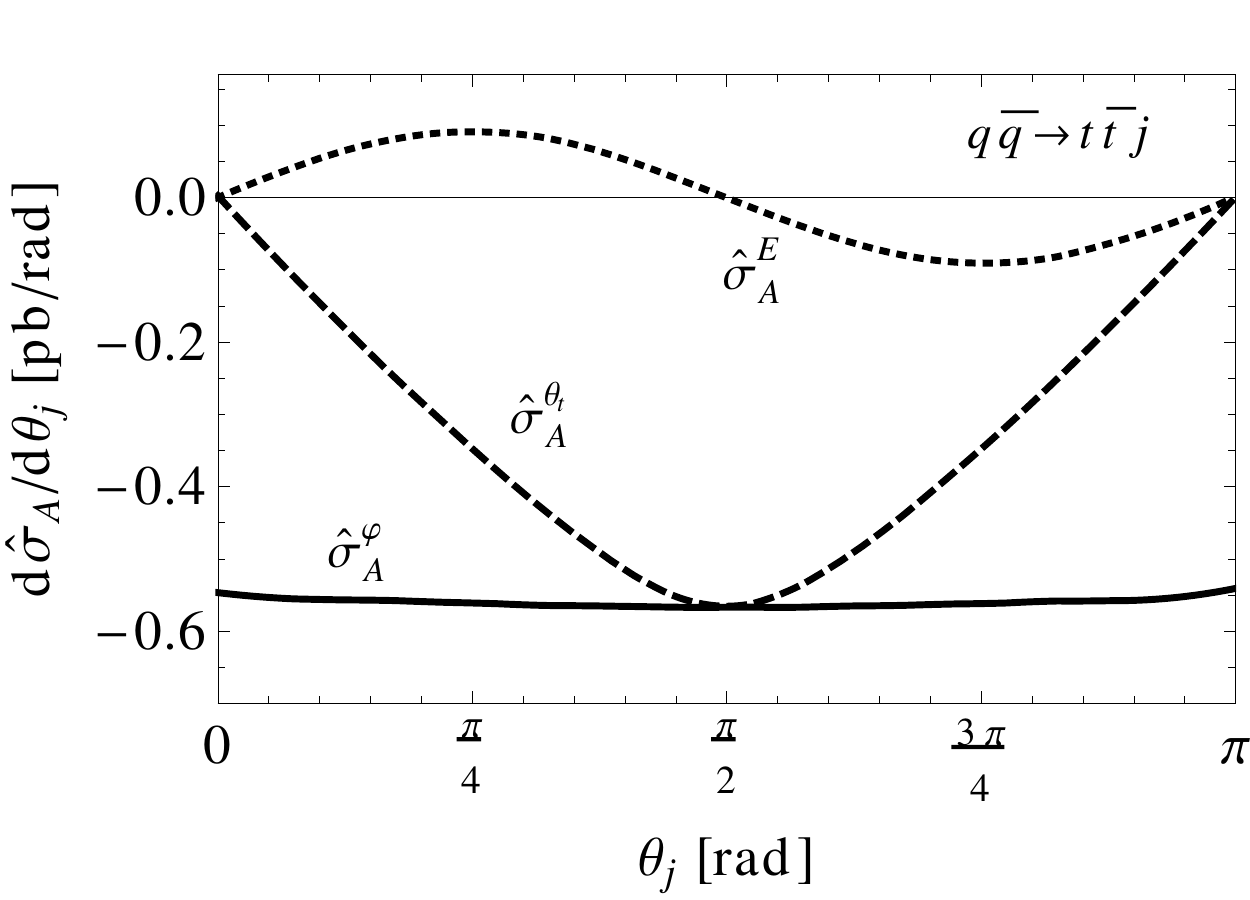}
\hspace*{0.5cm}
\includegraphics[height=5.5cm]{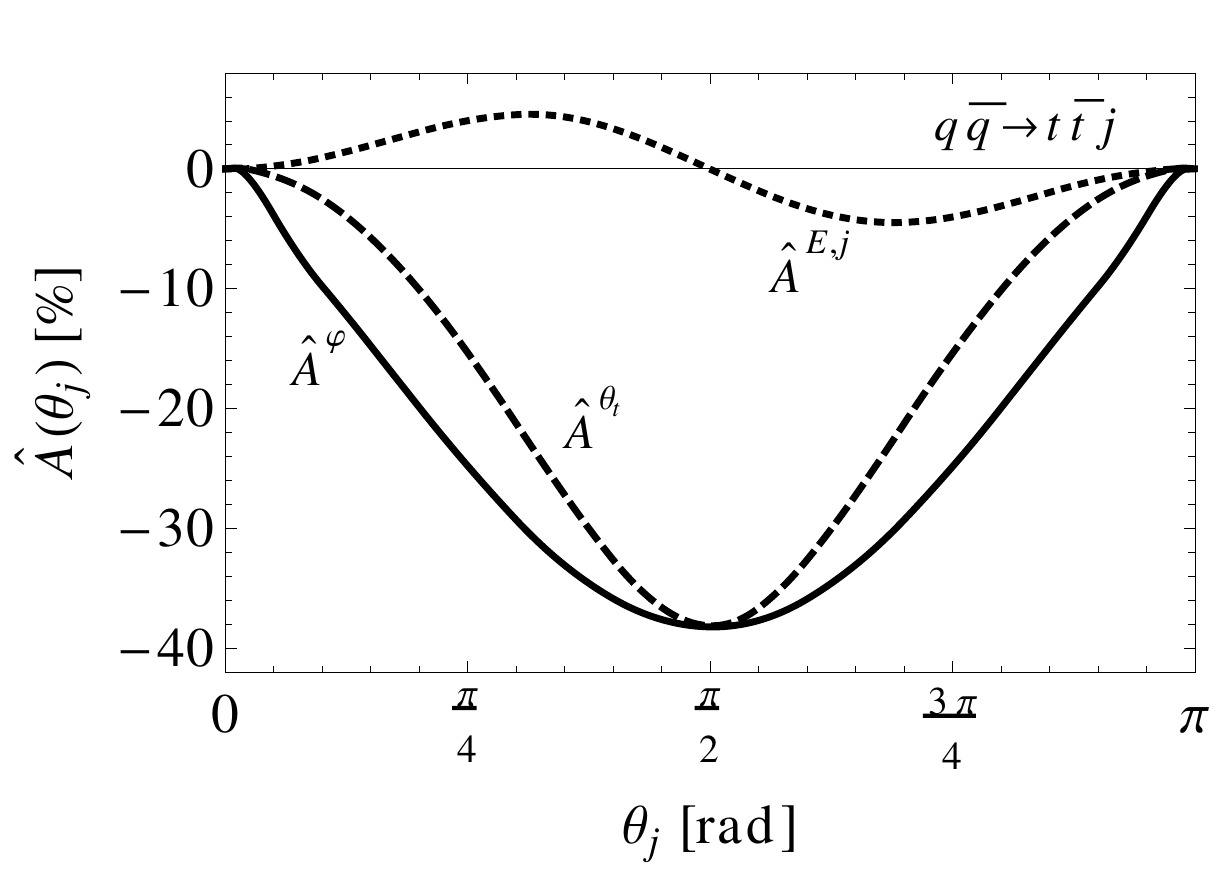}
\end{center}
\vspace*{-1cm}
\begin{center} 
  \parbox{15.5cm}{\caption{\label{fig:as-parton-qq} Partonic incline asymmetry (plain), energy asymmetry (dotted) and top-angle asymmetry (dashed) for the $q\bar q$ channel as functions of the jet scattering angle $\theta_j$ for $\sqrt{s} = 1\,\text{TeV}$ and $E_j \ge 20\,\text{GeV}$. Left: charge-asymmetric cross section $\text{d}\hat{\sigma}_A/\text{d}\theta_j$. Right: normalized asymmetry $\hat{A}(\theta_j)$. 
}}
\end{center}
\end{figure}

The top- and antitop-quark angles $\theta_t$ and $\theta_{\bar t}$ in the partonic CM frame can be expressed in terms of the variables introduced in (\ref{eq:parametrization}) as
\begin{eqnarray}\label{eq:ctphi}
\cos\theta_t & = & \sin\theta_j\cos\varphi\sin\xi + \cos\theta_j\cos\xi\,,\qquad\ \,\theta_t \in [0,\pi]\,,\\\nonumber 
\cos\theta_{\bar t} & = & \sin\theta_j\cos\varphi\sin\bar{\xi} + \cos\theta_j\cos\bar{\xi}\,,\qquad\ \,\theta_{\bar t} \in [0,\pi]\,. 
\end{eqnarray}
In Figure~\ref{fig:as-parton-qq}, left, we display the partonic incline asymmetry $\hat{\sigma}_A^{\varphi}$ (plain curve) and the top-angle asymmetry $\hat{\sigma}_A^{\theta_t}$ (dashed curve) for the $q\bar q$ channel as functions of $\theta_j$. We notice three {cha\-racte\-ristic} features: 1) The incline asymmetry $\hat{\sigma}_A^{\varphi}$ is largely independent from the jet direction. 2) The top-angle asymmetry $\hat{\sigma}_A^{\theta_t}$ reaches its maximum for central jet scattering, $\theta_j = \pi/2$, where it equals the incline asymmetry. 3) The top-angle asymmetry vanishes for $\theta_j = 0,\pi$. Let us discuss these features one by one.\\

1) The flatness of the distribution $\text{d}\hat{\sigma}_A^{\varphi}/\text{d}\theta_j$ has already been discussed in Section~\ref{subsec:aphi} in terms of the coefficients $\{\widetilde{N}_1\}$. It is due to the fact that the inclination between the top quark and the incoming quark is defined perpendicular to the jet direction. The impact of the jet kinematics on the direction of the top and antitop quarks, which tends to wash out the asymmetry, is thereby largely eliminated. The incline asymmetry is thus at its maximum over (almost) the entire range of $\theta_j$.\\

2) To demonstrate that the incline asymmetry  is equal to the top-angle (or forward-backward) asymmetry if the jet is emitted perpendicular to the beam axis, we consider both observables for $\theta_j = \pi/2$. In this limit, (\ref{eq:ctphi}) reduces to
\begin{eqnarray}
\cos\theta_t & = & \cos\varphi\sin\xi\,,\qquad\ \,\theta_j = \pi/2\,.  
\end{eqnarray}
The variables $\cos\theta_t$ and $\cos\varphi$ are now equal up to a factor $\sin\xi$, which is fixed by the energies $E_t$ and $E_{\bar t}$ (see (\ref{eq:cosxi})) and is positive over its range $\xi \in [0,\pi]$. A forward-scattered top quark ($\cos\theta_t > 0$) thus corresponds to a positive inclination $\cos\varphi > 0$, whereas a backward-scattered top ($\cos\theta_t < 0$) induces a negative inclination $\cos\varphi < 0$. The antitop quark is emitted in the respective other hemisphere. This implies that for $\theta_j = \pi/2$ the incline asymmetry and the forward-backward asymmetry are equal, once the integration over $\cos\varphi$ and $\cos\theta_t$ is performed,
\begin{eqnarray}
\hat{A}^{\varphi}(\theta_j = \pi/2) & = & \frac{\hat{\sigma}(\cos\varphi > 0) - \hat{\sigma}(\cos\varphi < 0)}{\hat{\sigma}_S}\Big \vert_{\theta_j=\pi/2}\\
 & = & \frac{\hat{\sigma}(\cos\theta_t > 0) - \hat{\sigma}(\cos\theta_t < 0)}{\hat{\sigma}_S}\Big \vert_{\theta_j=\pi/2} = \hat{A}^{\theta_t}(\theta_j = \pi/2)\,.\nonumber  
\end{eqnarray}
Away from the region of central jet emission, the top-angle asymmetry $\hat{A}^{\theta_t}$ is reduced. For $\theta_j \ne \pi/2$, the dependence of $\cos\theta_t$ and $\cos\theta_{\bar t}$ in (\ref{eq:ctphi}) on $\xi$ and $\bar{\xi}$ relaxes the correlation between the signs of $\cos\varphi$ and $\cos\theta_t-\cos\theta_{\bar t}$. For instance, events with a positive inclination $\cos\varphi > 0$ can correspond to constellations with either $\cos\theta_t > \cos\theta_{\bar t}$ or $\cos\theta_t < \cos\theta_{\bar t}$. The latter reduce the top-angle asymmetry $\hat{A}^{\theta_t}$ with respect to the incline asymmetry $\hat{A}^{\varphi}$.\\

3) The top-angle asymmetry vanishes if the jet is emitted along the beam axis, {cor\-re\-spon\-ding} to $\sin \theta_j = 0$. The top-quark scattering angle is now given by
\begin{eqnarray}
\cos\theta_t & = & \pm \cos\xi\,,\qquad\ \,\theta_j \,=\, 0,\pi\,. 
\end{eqnarray}
For a forward jet ($\theta_j = 0$), the variables $\cos\theta_t$ and $\cos\xi$ are equal; for a backward jet ($\theta_j = \pi$), they are opposite in sign. At this limit, the correlation between $\theta_t$ and $\varphi$ is lost. An inclination of the plane $(t,\bar t,j)$ with respect to $(q,\bar q,j)$ corresponds to an azimuthal rotation of the top quark around the beam axis, $\varphi = \phi_t$, which is orthogonal to the top-quark scattering angle $\theta_t$. The differential charge asymmetry takes the simple form
\begin{eqnarray}
\frac{\text{d}\hat{\sigma}_A(\theta_j = 0,\pi)}{\text{d}\varphi\,\text{d}E_t\,\text{d}E_{\bar t}} & \sim & - N_1(E_t,E_{\bar t})\,\cos\varphi\,.
\end{eqnarray}
Since $\cos\theta_t$ and $\cos\theta_{\bar t}$ can be expressed in terms of $E_t$ and $E_{\bar t}$ via $\cos\xi$ and $\cos\bar{\xi}$, the asymmetry $\hat{A}^{\theta_t}$ obviously vanishes for $\theta_j = 0,\pi$, once the integration over $\varphi \in [0,2\pi]$ is performed. In other words, the top-angle asymmetry vanishes for jet emission in the beam direction, because the top and antitop angles $\theta_t$ and $\theta_{\bar t}$ are orthogonal to the inclination angle $\varphi$.
\begin{figure}[!t]
\begin{center}
\includegraphics[height=5.5cm]{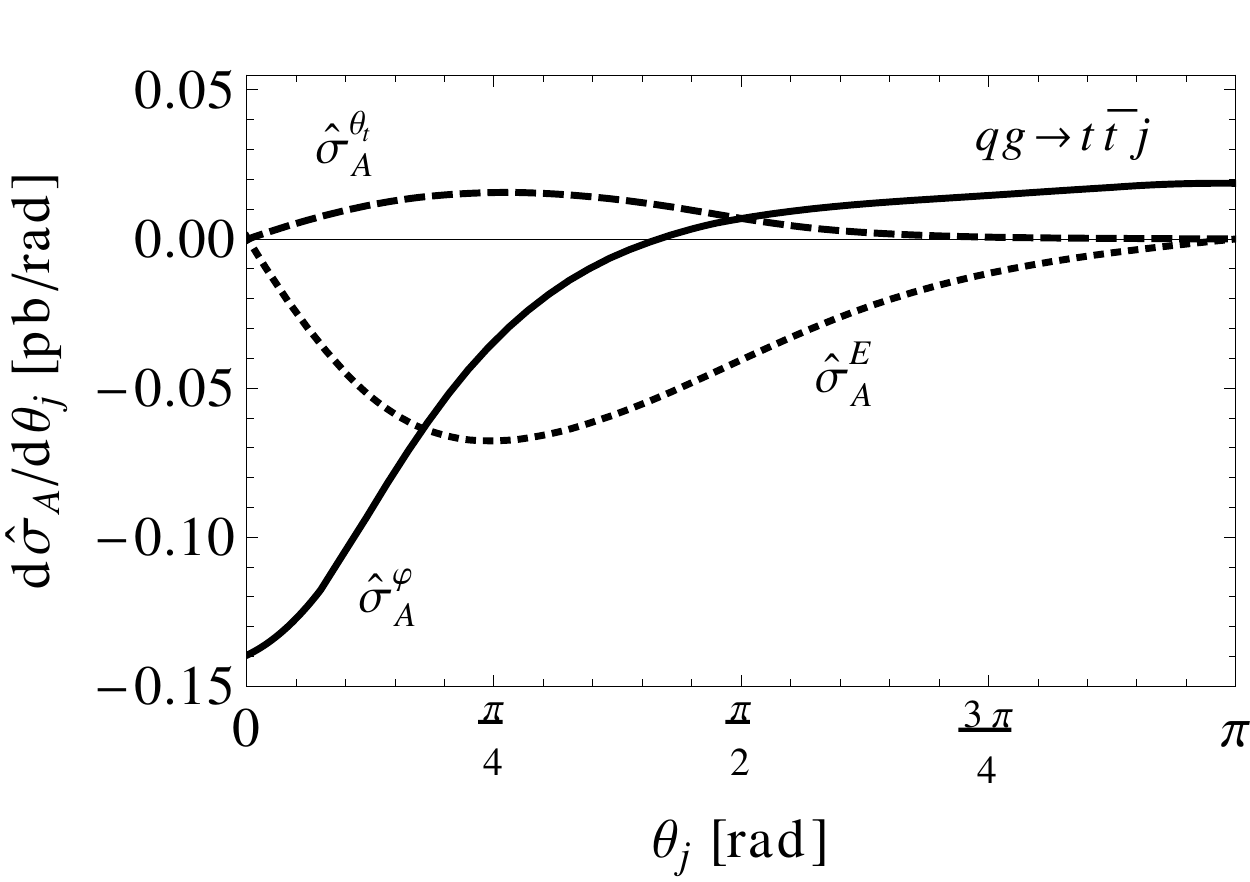}
\hspace*{0.5cm}
\includegraphics[height=5.5cm]{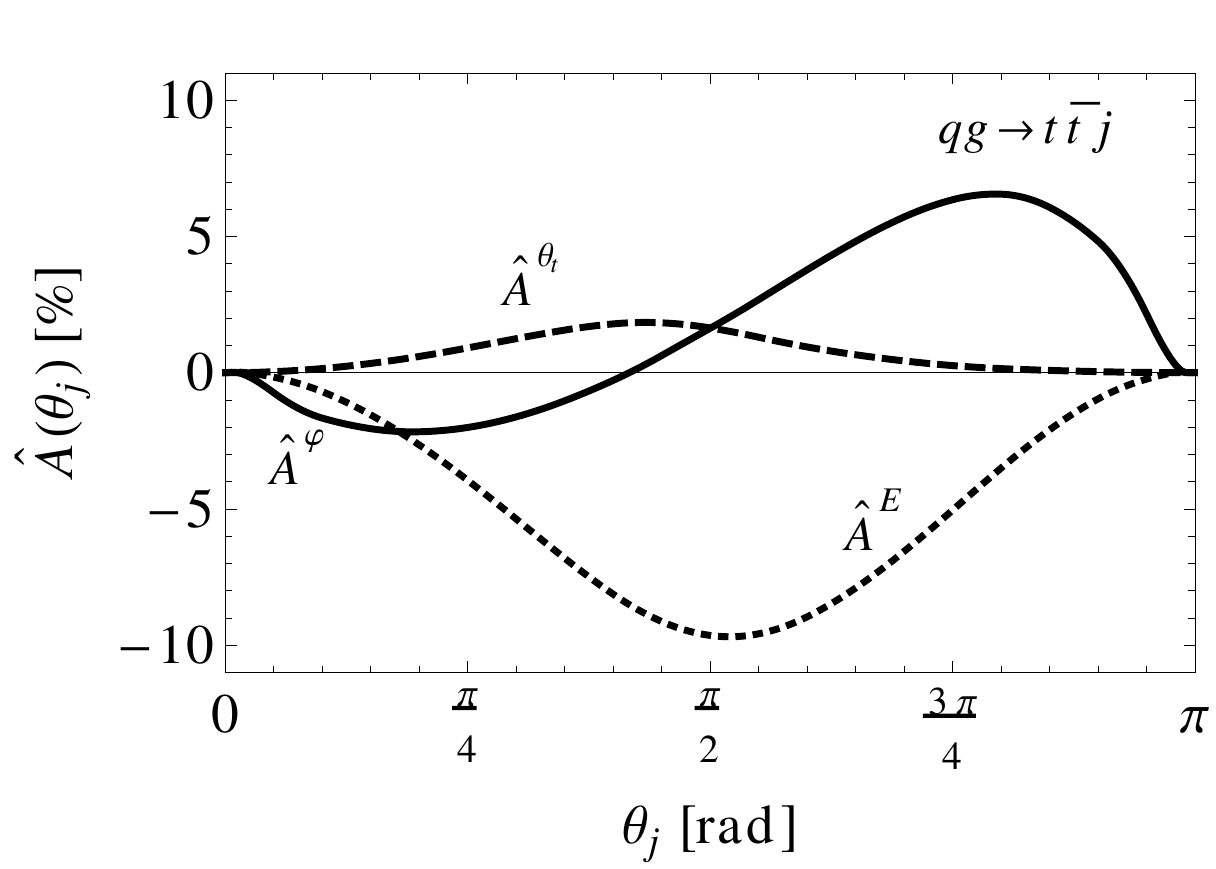}
\end{center}
\vspace*{-1cm}
\begin{center} 
  \parbox{15.5cm}{\caption{\label{fig:as-parton-qg} Partonic incline asymmetry (plain), energy asymmetry (dotted) and top-angle asymmetry (dashed) for the $qg$ channel as functions of the jet scattering angle $\theta_j$ for $\sqrt{s} = 1\,\text{TeV}$ and $E_j \ge 20\,\text{GeV}$. Left: charge-asymmetric cross section $\text{d}\hat{\sigma}_A/\text{d}\theta_j$. Right: normalized asymmetry $\hat{A}(\theta_j)$.}}
\end{center}
\end{figure}
\subsection{Summary: asymmetries in the $q\bar q$ and $qg$ channels}\label{subsec:asparton-summary}
Let us conclude our analysis by comparing the different asymmetries at the parton level in the $q\bar q$ and $qg$ channels. In Figures~\ref{fig:as-parton-qq} and \ref{fig:as-parton-qg}, the incline asymmetry (plain), energy asymmetry (dotted) and top-angle asymmetry (dashed) are shown as functions of the jet scattering angle $\theta_j$ in the partonic CM frame. To visualize the effect of the normalization to the symmetric cross section $\text{d}\hat{\sigma}_S/\text{d}\theta_j$, we display both the charge-asymmetric cross section $\text{d}\hat{\sigma}_A/\text{d}\theta_j$ (left panel) and the normalized asymmetry $\hat{A}(\theta_j)$ (right panel).

In the $q\bar q$ channel, all asymmetries are either symmetric or antisymmetric with respect to $\theta_j\rightarrow \pi - \theta_j$, because the jet distribution is forward-backward symmetric. Since the symmetric cross section is enhanced in the collinear region, the normalized asymmetries are suppressed for $\theta_j \approx 0,\pi$. The suppression is particularly strong for the incline asymmetry $\hat{A}^{\varphi}$, where the charge-antisymmetric cross section $\text{d}\hat{\sigma}^{\varphi}_A/\text{d}\theta_j$ is largely constant in $\theta_j$. For the energy asymmetry $\hat{A}^{E,j}$ and the top-angle asymmetry $\hat{A}^{\theta_t}$, where per definition $\text{d}\hat{\sigma}^{E,j}_A/\text{d}\theta_j$ and $\text{d}\hat{\sigma}^{\theta_t}_A/\text{d}\theta_j$ decrease for $\theta_j \rightarrow 0,\pi$, the shape distortion by the normalization is less significant. The energy asymmetry is much smaller than the incline asymmetry and the top-angle asymmetry. Its maximum amounts to 
$|\hat{A}^{E,j}| = 5\,\%$ for $\theta_j = 3\pi/10$ and $7\pi/10$ for $\sqrt{s}=1\,\text{TeV}$ and $E_j\ge 20\,\text{GeV}$.  
As has been discussed in Section~\ref{subsec:afb}, the incline and top-angle asymmetries are equal at their maximum at $\theta_j = \pi/2$, where $|\hat{A}^{\varphi}| = |\hat{A}^{\theta_t}| = 38\,\%$.  
For $\theta_j \neq \pi/2$, $\hat{A}^{\varphi}$ is enhanced with respect to $\hat{A}^{\theta_t}$ over the entire range of the jet scattering angle. The incline asymmetry is therefore superior to the top-angle asymmetry as a probe of the charge asymmetry in $q\bar q\rightarrow t\bar t j$. The maximum of the asymmetries at the parton level serves as a guideline for the upper bound on the expected hadronic observables. The size of the effective observables at the hadron level can be significantly reduced, dependent on the amount of charge-symmetric background and on the cuts imposed on the signal.

In the $qg$ channel, the jet distribution of the asymmetries is neither symmetric nor antisymmetric, due to the asymmetric kinematics of the process $qg\rightarrow t\bar tq$. Since the quark-jet is preferentially emitted in the direction of the incident quark, the antisymmetric cross section $\text{d}\hat{\sigma}_A/\text{d}\theta_j$ is enhanced, but finite, for small angles $\theta_j$. The enhancement of the symmetric cross section $\text{d}\hat{\sigma}_S/\text{d}\theta_j$ for $\theta_j\rightarrow 0$, however, is stronger due to its collinear divergence. The maximum of the normalized asymmetry $\hat{A}(\theta_j)$ is thus reached for larger jet angles $\theta_j$. The top-angle asymmetry in the $qg$ channel is tiny. Its maximum amounts to no more 
than $|\hat{A}^{\theta_t}| = 2\,\%$ at $\theta_j = 9\pi/20$.  
The incline asymmetry exhibits a larger maximum of $|\hat{A}^{\varphi}| = 7\,\%$ at $\theta_j = 4\pi/5$.  
However, since the inclination angle $\varphi$ in the $qg$ CM frame is defined with respect to the quark-jet rather than the gluon, $\hat{A}^{\varphi}$ changes sign around $\theta_j = 2\pi/5$, such that the integrated incline asymmetry is much smaller. The energy asymmetry is the appropriate observable of the charge asymmetry in the $qg$ CM frame. Its maximum is sizeable, $|\hat{A}^E| = 10\,\%$, and occurs for jet angles close to $\theta_j = \pi/2$. The latter feature is particularly useful, because  (as for the incline asymmetry) it allows one to cut out the collinear divergence of the normalization $\hat{\sigma}_S$ by focusing on the region of central jet emission. We elaborate on the origin and size of the different asymmetries in the $q\bar q$ and $qg$ channels in Appendix~\ref{app:boosted}.

In the following sections, we will mostly focus on the asymmetries $\hat{A}^{\varphi}$ for the $q\bar q$ channel and $\hat{A}^{E}$ for the $qg$ channel. In order to construct robust observables for hadron colliders, it is crucial to control the jet energy and angular distributions of the charge asymmetries. The purpose of our analysis in QCD at LO is to show the kinematical behavior of the asymmetries at a qualitative level. A numerically precise prediction requires QCD calculations beyond LO and the resummation of the logarithmic infrared and collinear divergences. This is particularly important, as hadronic observables will depend on experimental cuts on the jet energy and transverse momentum, which regularize the divergent regions of the phase space. A stronger cut on the jet scattering angle that projects on the central region around $\theta_j \approx \pi/2$ simultaneously enhances the asymmetries and ensures collinear-safe observables.

\begin{figure}[!t]
\begin{center}
\includegraphics[height=5cm]{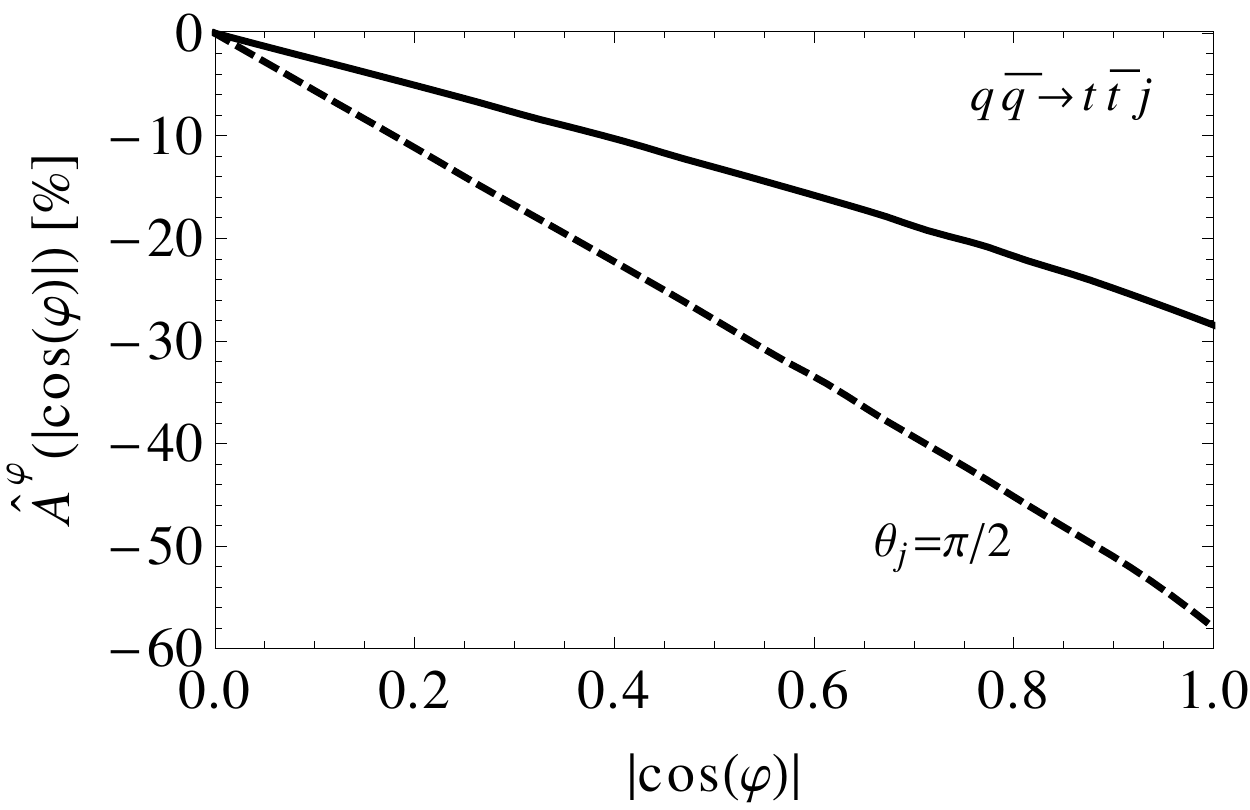}
\hspace*{0.5cm}
\includegraphics[height=5cm]{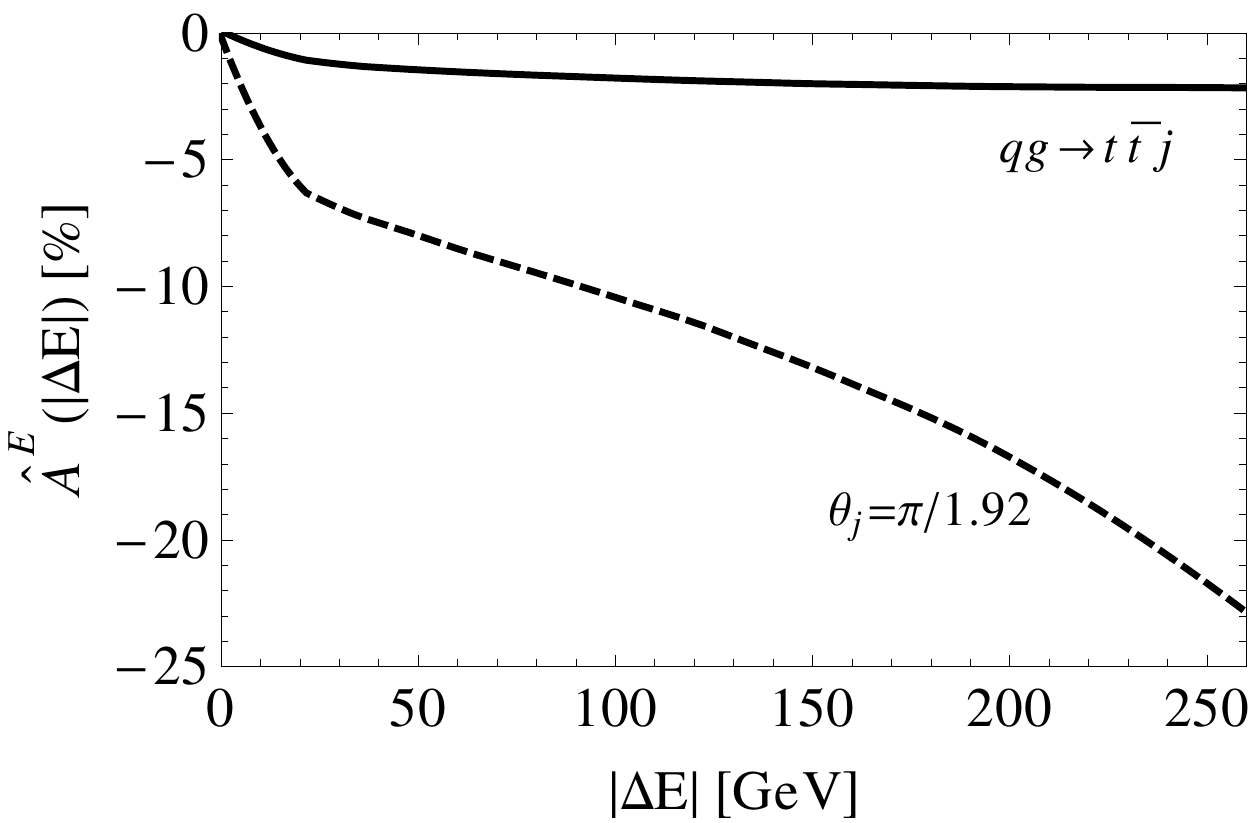}
\end{center}
\vspace*{-1cm}
\begin{center} 
  \parbox{15.5cm}{\caption{\label{fig:as-parton-decp} Partonic distributions of charge asymmetries for $\sqrt{s} = 1\,\text{TeV}$ and $p_T^j \ge 20\,\text{GeV}$. Left: incline asymmetry $\hat{A}^{\varphi}(|\cos\varphi|)$ for the $q\bar q$ channel as a function of the inclination angle $|\cos\varphi|$ integrated over $\theta_j$ (plain) and for $\theta_j = \pi/2$ (dashed). Right: energy asymmetry $\hat{A}^E(|\Delta E|)$ for the $qg$ channel as a function of the energy difference $|\Delta E| = |E_t - E_{\bar t}|$ integrated over $\theta_j$ (plain) and for $\theta_j = \pi/1.92$ (dashed).}}
\end{center}
\end{figure}

At the LHC, further cuts on kinematic variables are needed to increase the sensitivity to the asymmetries. To this end, we investigate the dependence of $\hat{A}^{\varphi}$ and $\hat{A}^{E}$ on the respective defining variables $\cos\varphi$ and $\Delta E$. In Figure~\ref{fig:as-parton-decp}, we show the partonic incline asymmetry $\hat{A}^{\varphi}$ in $q\bar q\rightarrow t\bar t j$ (left panel) and the energy asymmetry $\hat{A}^E$ in $qg\rightarrow t\bar t j$ (right panel) as functions of $|\cos\varphi|$ and $|\Delta E|$. A cut on the transverse momentum of the jet, $p_T^j \ge 20\,\text{GeV}$, has been applied to regularize the infrared and collinear divergences. Plain curves correspond to observables integrated over the jet distribution, whereas dashed curves show the observables at their maxima, $\hat{A}^{\varphi}(\theta_j = \pi/2)$ and $\hat{A}^E(\theta_j = \pi/1.92)$.  
The asymmetries increase with $|\cos\varphi|$ and $|\Delta E|$, respectively. This effect is maximized, if the jet is emitted perpendicular to the beam axis in the partonic CM frame, namely at $\theta_j\approx \pi/2$. In this region, the cross section $\hat{\sigma}_S$ does not exhibit a collinear enhancement, such that the sensitivity of the normalized asymmetries $\hat{A} = \hat{\sigma}_A/\hat{\sigma}_S$ to the properties of $\hat{\sigma}_A$ is higher than in the collinear regime around $\theta_j \approx 0,\pi$.

The incline asymmetry $\hat{A}^{\varphi}$ in the $q\bar q$ channel increases almost linearly with $|\cos\varphi|$, reflecting the fact that the differential asymmetric cross section $\text{d}\hat{\sigma}_A$ is proportional to $\cos\varphi$ (see (\ref{eq:sigmaa})). The maximum of $\hat{A}^{\varphi}$ is reached for $|\cos\varphi|\approx 1$, where the planes $(t,\bar t,j)$ and $(q,\bar q,j)$ coincide. This is also the region with the highest production rate of $t\bar t + j$ events. A lower cut on $|\cos\varphi|$ thus increases the asymmetry without too strong a reduction of the cross section. In hadron collisions at the Tevatron and the LHC, the total cross section for $t\bar t + j$ production is dominated by processes with a partonic CM energy of $\sqrt{s}\approx 500-600\,\text{GeV}$. Since the dependence of $\hat{A}^{\varphi}$ on $\sqrt{s}$ is mild, the partonic results for $\sqrt{s} = 1\,\text{TeV}$ from Figure~\ref{fig:as-parton-decp}, left, already give a rough estimate of the incline asymmetry at hadron level. In particular, it can be observed that the incline asymmetry at the LHC can be significantly enhanced by suitably combined cuts on $\theta_j$ and $|\cos\varphi|$.

The situation is different for the energy asymmetry $\hat{A}^E$ in the $qg$ channel. For kinematical reasons, a large top-antitop energy difference $|\Delta E|$ implies a lower bound on the jet energy $E_j$ (see Figure~\ref{fig:N2phiDE}, left). A lower cut on $|\Delta E|$ thus enhances the energy asymmetry, but comes along with a significant reduction of the cross section. As is apparent in Figure~\ref{fig:as-parton-decp}, right, the increase of $\hat{A}^E$ with $|\Delta E|$ is much more pronounced for central jet emission around $\theta_j\approx \pi/2$. A cut on the jet angle thus will be important to raise the energy asymmetry to an observable level at the LHC.

\section{Asymmetries at hadron colliders}\label{sec:colliders}
With these partonic investigations at hand, we construct observables for the Tevatron and LHC experiments that probe the charge asymmetry in both the $q\bar q$ and the $qg$ channel. At the hadron level, the partonic charge asymmetries introduced in Section~\ref{sec:asymmetry} need to be adjusted according to the proton-antiproton and proton-proton initial states, respectively. The hadronic cross sections $\sigma_{S/A}$ are obtained from the partonic cross sections $\hat{\sigma}_{S/A}^{p_1 p_2}$ by a convolution with the parton distribution functions (PDFs) $f_{p_{1,2}/P_{1,2}}(x_{1,2},\mu_f)$ at the QCD factorization scale $\mu_f$ and the sum over all partonic initial states $p_1 p_2$,
\begin{eqnarray}\label{eq:hadronicxsec}
\sigma_{S/A}(S) = \sum_{p_1 p_2}\iint \text{d} x_1 \text{d}x_2 \,f_{p_1/P_1}(x_1,\mu_f)\,f_{p_2/P_2}(x_2,\mu_f)\,\hat{\sigma}_{S/A}^{p_1 p_2}(s = x_1 x_2 S,\mu_f)\,.  
\end{eqnarray}
Here $x_{1,2}$ denotes the momentum fraction of the parton $p_{1,2}$ inside the nucleon $P_{1,2}$, and $\sqrt{s}$ and $\sqrt{S}$ are the partonic and hadronic CM energies. For the quark-antiquark initial state $p_1 p_2 = q\bar q$, which dominates the top-quark pair production rate at the Tevatron, the differential partonic charge-asymmetric cross section $\text{d}\hat{\sigma}_A(q\bar q\rightarrow t\bar t j)/\text{d}\varphi \text{d}\theta_j \text{d}E_t \text{d}E_{\bar t}$ has been given in (\ref{eq:sigmaa}). At the LHC, the quark-gluon initial state $p_1 p_2 = qg$ will also be important. The boost of the partonic CM frame with respect to the laboratory frame will be expressed in terms of the rapidity of the top-antitop-jet system in the laboratory frame, $y_{t\bar t j} = \ln(x_1/x_2)/2$.

In our numerical analysis, the factorization scale is set equal to the top-quark mass, $\mu_f = m_t = 173.2\,\text{GeV}$. All calculations are performed at LO QCD, using CTEQ6L1 PDFs \cite{Pumplin:2002vw} and the corresponding value of the strong coupling constant, $\alpha_s^{\text{LO}}(m_t) = 0.1180$. The phase-space integration is performed numerically by means of the Vegas Monte Carlo algorithm implemented in \cite{Hahn:2004fe,Galassi}. Unless stated otherwise, we will apply typical experimental cuts on the jet's transverse momentum, $p_T^j$, and the jet rapidity in the laboratory frame, $y_j$. For the Tevatron analysis, we use $p_T^j \ge 20\,\text{GeV}$, $|y_j| \le 2$, and for the LHC we apply $p_T^j \ge 25\,\text{GeV}$, $|y_j| \le 2.5$, subsequently referred to as ``detector cuts''. Cuts on the jet angle $\theta_j$ will be expressed in terms of the partonic jet rapidity,
\begin{eqnarray}
 \hat{y}_j = \frac{1}{2}\log\left(\frac{1+\cos\theta_j}{1-\cos\theta_j}\right) = y_j - y_{t\bar t j}\,.
\end{eqnarray}
The jet scattering angle in the parton frame can thus be accessed by measuring the difference of the rapidities $y_j$ and $y_{t\bar tj}$ in the laboratory frame.

\subsection{Incline asymmetry at the Tevatron}\label{subsec:iatevatron}
\begin{figure}[!t]
\begin{center}
\includegraphics[height=5.5cm]{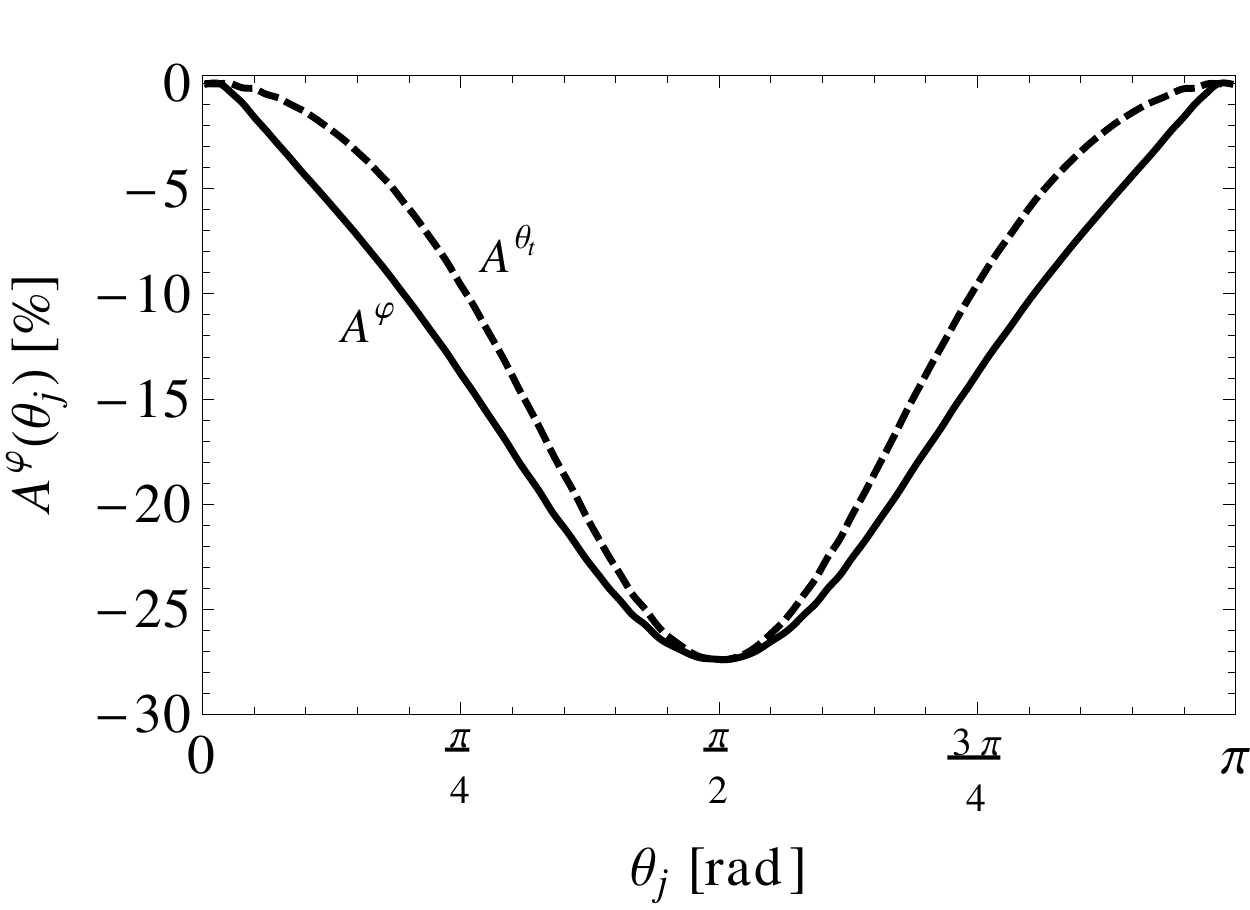}
\hspace*{0.5cm}
\includegraphics[height=5.5cm]{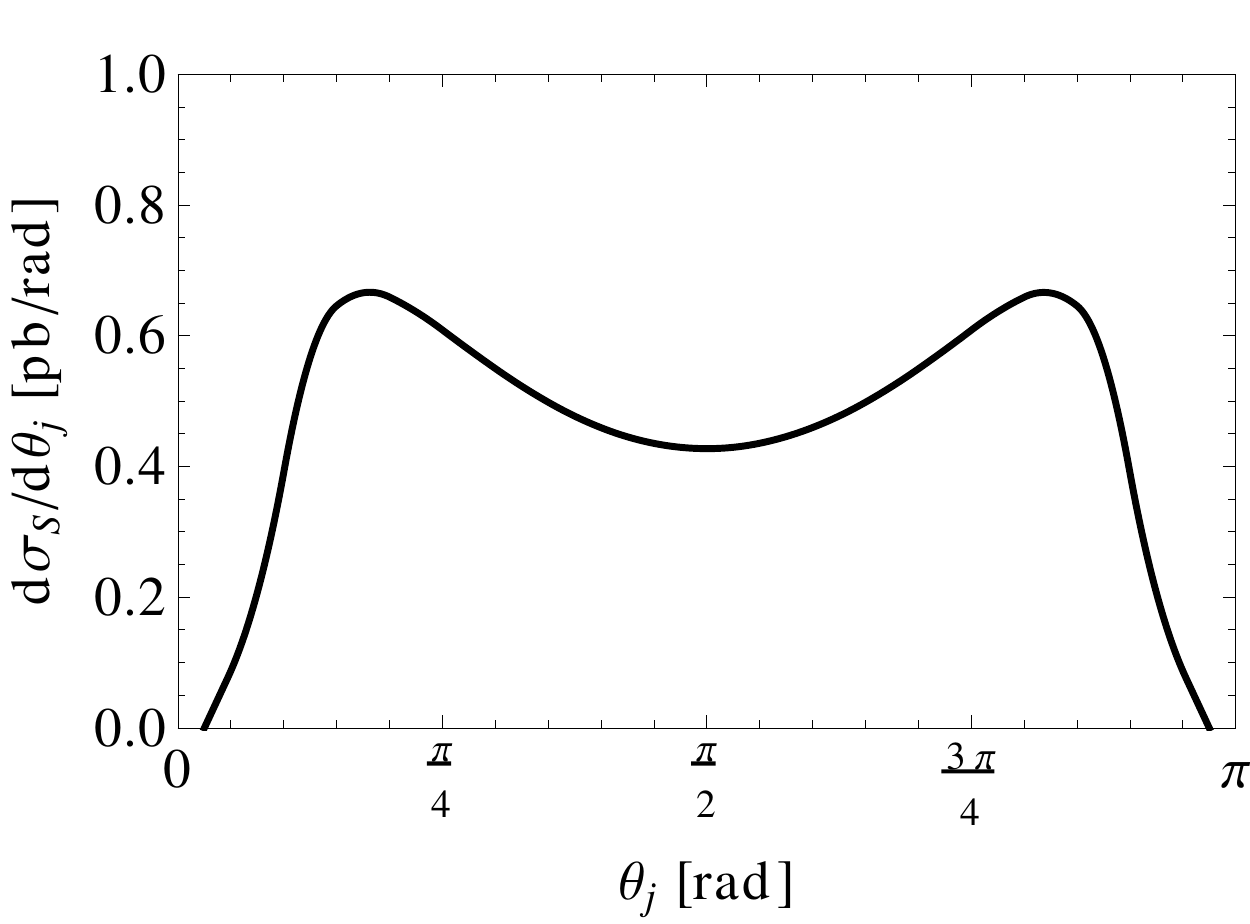}
\end{center}
\vspace*{-1cm}
\begin{center} 
  \parbox{15.5cm}{\caption{\label{fig:aphi-tev}Left: incline asymmetry $A^{\varphi}$ (plain) and top-angle asymmetry $A^{\theta_t}$ (dashed) as functions of the jet scattering angle $\theta_j$ at the Tevatron. Detector cuts of $p_T^j \ge 20\,\text{GeV}$ and $|y_j| \le 2$ are applied. Right: differential cross section $\text{d}\sigma_S/\text{d}\theta_j$. 
}}
\end{center}
\end{figure}

Proton-antiproton collisions at the Tevatron primarily proceed through the partonic $q\bar q$ channel. For a CM energy of $\sqrt{S} = 1.96\,\text{TeV}$ and detector cuts $p_T^j \ge 20\,\text{GeV}$ and $|y_j| \le 2$, the contributions of the partonic states to the total cross section of $t\bar t + j$ production amount to $84\%$ ($q\bar q$), $9\%$ ($qg + \bar q g$) and $7\%$ ($gg$)  
at LO QCD. The feature of the $q\bar q$ initial state being antisymmetric under charge conjugation is thus largely preserved in proton-antiproton collisions at the hadron level. The hadronic incline asymmetry $A^{\varphi}$ can be straightforwardly derived from the partonic asymmetry $\hat{A}^{\varphi}$ using (\ref{eq:aphi}) and (\ref{eq:hadronicxsec}),
\begin{eqnarray}
A^{\varphi} & = & \frac{\sigma_A^{\varphi}}{\sigma_S} = \frac{\int_{0}^{\pi} \text{d}\theta_j\,\text{d}\sigma_A^{\varphi}}{\sigma_S}\,.
\end{eqnarray}
In Figure~\ref{fig:aphi-tev}, we display the incline asymmetry $A^{\varphi}(\theta_j)$ (left panel, plain curve) and the symmetric cross section $\text{d}\sigma_S/\text{d}\theta_j$ (right panel) at the Tevatron as functions of the jet angle $\theta_j$.
The {ma\-xi\-mum} of $A^{\varphi} = -27.4\,\%$  
is obtained for jets emitted perpendicular to the beam axis in the partonic CM frame, namely for $\theta_j = \pi/2$. 
Recall from Section~\ref{subsec:afb} that the partonic asymmetry $\hat{\sigma}_A^{\varphi}$ is largely independent from $\theta_j$ (Figure~\ref{fig:as-parton-qq}, left). The drop-off of the normalized incline asymmetry at the edges of the spectrum is mainly due to the collinear enhancement of the symmetric cross section $\text{d}\sigma_S/\text{d}\theta_j$ for $\theta_j\rightarrow 0,\pi$. A cut on $p_T^j = E_j\sin\theta_j$ implies a minimal jet energy of $E_j = p_T^j$ for $\theta_j= \pi/2$ and of $E_j\gg p_T^j$ for $\theta_j\to 0,\pi$, suppressing the infrared enhancement of the cross section at the edges of the jet distribution. However, as the $p_T^j$ cut affects both $\sigma_A^{\varphi}$ and $\sigma_S$ in a similar way, the effect on the normalized incline asymmetry $A^{\varphi}$ is small.

For comparison, we also show the top-angle asymmetry $A^{\theta_t}$ (Figure~\ref{fig:aphi-tev}, left, dashed),
\begin{eqnarray}\label{eq:afbt}
A^{\theta_t} & = & \frac{\sigma_A^{\theta_t}}{\sigma_S} = \frac{\sigma(\Delta y > 0) - \sigma(\Delta y < 0)}{\sigma(\Delta y > 0) + \sigma(\Delta y < 0)} = A^y\,, 
\end{eqnarray}
which corresponds to the asymmetry $A^y$ in terms of the top-antitop rapidity difference $\Delta y = y_t - y_{\bar t}$. Independently from detector cuts and normalization effects, the definition of $\sigma_A^{\theta_t}$ implies a drop-off of the asymmetry for $\theta_j \rightarrow 0,\pi$ (see Figure~\ref{fig:as-parton-qq}, left). Therefore, $A^{\varphi}$ is always larger than $A^{\theta_t}$ for any fixed jet angle $\theta_j\ne \pi/2$. Integrated over the spectrum, the total asymmetries amount to 
$A^{\varphi} = -15.6\,\%$ 
and 
$A^{\theta_t} = -12.6\,\%$ 
for a cross section of 
$\sigma_S = 1.42\,\text{pb}$. 
The enhancement amounts to $A^{\varphi}/A^{\theta_t} = 1.24$ for the applied detector cuts. In the region of central jet emission, the enhancement is less pronounced.\\
\begin{table}[!b]
\begin{center}
\begin{tabular}{|c||c|c|c|c||c|c|}
\hline 
$(p_T^j)_{\text{min}}\,[\text{GeV}]$ & $20$ & $20$ & $20$ & $20$ & $15$ & $25$ \tabularnewline
$|\hat{y}_{j}|_{\text{max}}$ & --- & $1$ & --- & $1$ & --- & --- \tabularnewline
$|\cos\varphi|_{\text{min}}$ & --- & --- & $0.6$ & $0.6$ & --- & --- \tabularnewline
\hline
\hline
$A^{\varphi}\,\,[\%]$ & $-15.6$ & $-20.6$ & $-20.9$ & $-27.5$ & $-15.5$ & $-15.6$ \tabularnewline
\hline 
$\sigma_{S}\,\,[\text{pb}]$ & $1.42$ & $0.87$ & $0.86$ & $0.53$ & $1.83$ & $1.14$ \tabularnewline
\hline
\hline
$\mathcal{S}(10\,\text{fb}^{-1})$ & $4.2$ & $4.3$ & $4.3$ & $4.5$ & $4.7$ & $3.7$ \tabularnewline
\hline
\end{tabular}
\end{center}
\begin{center} 
  \parbox{15.5cm}{\caption{\label{tab:tev_cuts} Incline asymmetry $A^{\varphi}$, cross section $\sigma_S$ and statistical significance $\mathcal{S} = |A^{\varphi}|/\delta A^{\varphi}$ at the Tevatron. Detector cuts $|y_j| \le 2$ and $p_T^j \ge 20\,\text{GeV}$ (altered in the last two columns) have been applied, as well as additional cuts on the partonic jet rapidity $\hat{y}_j$ and the inclination angle $\varphi$.}}
\end{center}
\end{table}
In Table~\ref{tab:tev_cuts}, we give numerical values for the integrated incline charge asymmetry $A^{\varphi}$ and the corresponding cross section $\sigma_S$ and their dependence on various cuts. The normalized asymmetry is stable against moderate variations of the detector cut on $p_T^j$ (last two columns). As suggested by Figure~\ref{fig:aphi-tev}, the dependence of both $A^{\varphi}$ and $\sigma_S$ on the jet direction is strong. Since the asymmetry is maximized for jets emitted perpendicular to the beam axis, an additional upper bound on the absolute jet rapidity $|\hat{y}_j|$ in the partonic CM frame (third column) increases the integrated asymmetry. The price to pay is a significant reduction of the total cross section. A similar effect is obtained by a suitable cut on the inclination angle $\varphi$ (fourth column). Requiring simultaneously a central jet and a small inclination between the initial- and final-state planes (fifth column), enhances $A^{\varphi}$  more efficiently than a stronger cut on either $|\hat{y}_j|$ or $\varphi$ yielding the same cross section.

In practice, the significance of an observation of the incline asymmetry at the Tevatron will be limited by the low $t\bar t + j$ production rate. In the last row of Table~\ref{tab:tev_cuts},
 we display the statistical significance $\mathcal{S} = |A^{\varphi}|/\delta A^{\varphi}$ of the integrated asymmetry with $\delta A^{\varphi} = 1/\sqrt{N}$. The expected number of $t\bar t + j$ events amounts to $N = 711$ 
 for our default detector cuts, assuming an integrated luminosity of $\mathcal{L}=10\,\text{fb}^{-1}$ and an experimental efficiency of $5\%$. For the cuts displayed in Table~\ref{tab:tev_cuts}, we obtain a statistical significance of $\mathcal{S}\approx 4-5$.  
The significance is maximized for loose cuts, because of the statistical limitations. A detailed experimental analysis is required to optimize the set of cuts, in order to maximize $A^{\varphi}$ while maintaining most of the cross section $\sigma_S$. Despite the limited statistics, the potential to measure the incline asymmetry at the Tevatron is substantial and should be pursued as an important test of QCD.

We close this subsection with remarks on the energy asymmetry in $p\bar p$ collisions at the Tevatron. The dominant contribution to the energy asymmetry stems from the antisymmetric $q\bar q$ initial state. We therefore can apply the parton-level definition of $\hat{A}^{E,j}$ in (\ref{eq:energy-asymmetry-qqb}) at the hadron level by merely integrating over the PDFs to obtain $A^{E,j} = \sigma_A^{E,j}/\sigma_S$. The impact of the $qg$ initial state on $A^{E,j}$ is numerically completely negligible. As we have observed at the parton level (see Figure~\ref{fig:as-parton-qq}, right), the energy asymmetry $\hat{A}^{E,j}$ in the $q\bar q$ channel is considerably smaller than the incline asymmetry $\hat{A}^{\varphi}$. At the Tevatron, the total energy asymmetry amounts to no more than $A^{E,j} = 0.9\%$. 
 Given the low production rate of $t\bar t + j$ events, it is not possible to significantly enhance the energy asymmetry by cuts. An observation of $A^{E,j}$ at the Tevatron is thus out of reach.

\subsection{Incline asymmetry at the LHC}\label{subsec:ialhc}
Due to the high CM energy in proton-proton collisions at the LHC, $t\bar t$ pairs are frequently produced in association with a hard jet. For $\sqrt{S} = 8\,\text{TeV}$ and typical detector cuts $p_T^j \ge 25\,\text{GeV}$ and $|y_j| \le 2.5$, the cross section for the production of a $t\bar t + j$ state amounts to $\sigma_S = 97.5\,\text{pb}$  
 at LO QCD, more than $50\%$ 
 of the total cross section for inclusive $t\bar t$ production. Compared to the Tevatron, two main differences affect the definition of a charge asymmetry at the LHC. Firstly, the initial proton-proton state is charge-symmetric. A non-vanishing charge asymmetry thus requires a reference frame other than the direction of the incoming hadrons. Secondly, the distribution of the partonic states from proton-proton collisions at the LHC differs significantly from the situation at the Tevatron. For $\sqrt{S} = 8\,\text{TeV}$ and applied detector cuts, the total cross section at LO is composed of the partonic contributions to $65.6\%$ ($gg$), $26.7\%$ ($qg + \bar q g$) and $7.7\%$ ($q\bar q$). 
To observe a charge asymmetry, the large charge-symmetric $gg$ background needs to be suppressed as far as possible by suitable cuts. Whether the asymmetry is then dominated by $q\bar q$ or $qg$ contributions depends on the respective observable and on the applied cuts. As we will show, it is possible to disentangle $q\bar q$ contributions from $qg$ contributions by accessing the charge asymmetry through different observables. As a suitable set of such observables, we propose an incline asymmetry to probe the $q\bar q$ channel and an energy asymmetry to probe the $qg$ channel.

\begin{figure}[!t]
\begin{center}
\hspace*{-0.2cm}
\raisebox{0.19cm}{\includegraphics[height=5cm]{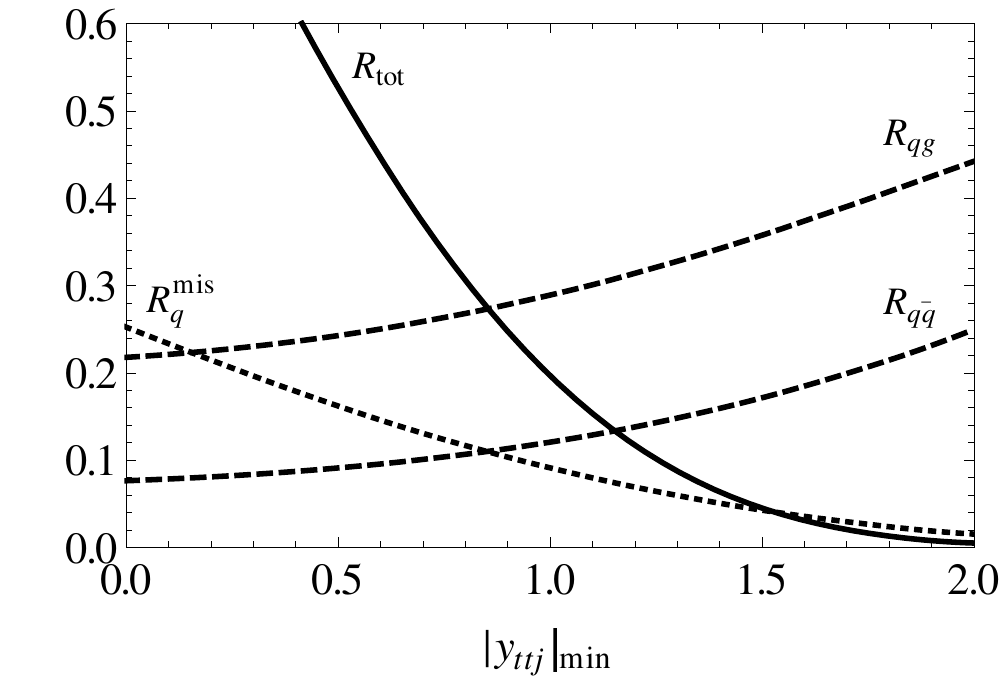}}
\hspace*{0.3cm}
\includegraphics[height=5.49cm]{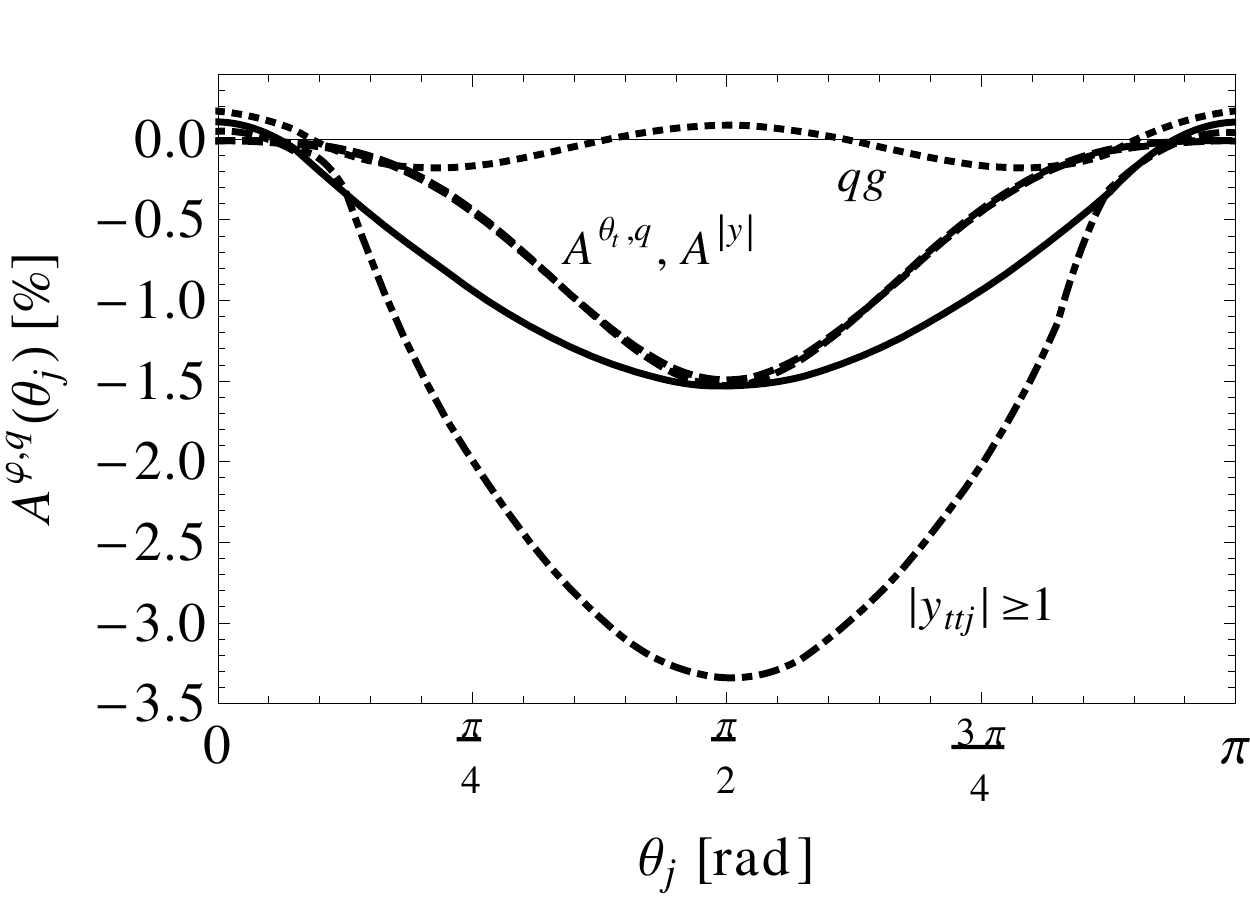}
\end{center}
\vspace*{-1cm}
\begin{center} 
  \parbox{15.5cm}{\caption{\label{fig:aphi-lhc}Left: effects of a cut on the final-state boost, $|y_{t\bar tj}|_{\text{min}}$, in $t\bar t + j$ production at LHC8. Shown are the reduction of the total cross section, $R_{\text{tot}}$ (plain), the fractions of $q\bar q + \bar q q$ and $qg + gq$ partonic states in the total cross section, $R_{q\bar q}$ and $R_{qg}$ (dashed), and the fraction of misinterpreted quarks in $q\bar q$ states, $R_q^{\text{mis}}$ (dotted). Right: incline asymmetry $A^{\varphi,q}(\theta_j)$ at LHC8 without (plain) and with a boost cut of $|y_{t\bar tj}| \ge 1$ (dot-dashed), and the $qg$ pollution in $A^{\varphi,q}(\theta_j)$ (dotted). For comparison are shown the top-angle asymmetry $A^{\theta_t,q}(\theta_j)$ (dashed) and the rapidity asymmetry $A^{|y|}(\theta_j)$ (dashed, overlapping with $A^{\theta_t,q}(\theta_j)$). Detector cuts of $p_T^j \ge 25\,\text{GeV}$ and $|y_j| \le 2.5$ are applied.}}
\end{center}
\end{figure}

The partonic incline asymmetry $\hat{A}^{\varphi}$ in $q\bar q\rightarrow t\bar t j$ has been defined in (\ref{eq:aphi}). The definition of the inclination angle $\varphi$ in (\ref{eq:angles}) involves the momentum $\vec{k}_1$ of the incident quark $q$ in the partonic initial state $p_1 p_2$. We will guess the direction of the quark by exploiting the fact that valence quarks carry most of the longitudinal momentum of the incident proton. The final state is thus boosted in the direction of the valence quark, encoded in the rapidity $y_{t\bar t j} = \ln(x_1/x_2)/2$. 
If $y_{t\bar tj} > 0$, the quark is supposed to stem from the proton $P_1$ ($q=p_1$); if $y_{t\bar tj} < 0$, it is supposed to originate from the proton $P_2$ ($q=p_2$). The quark direction serves as a reference frame to define the incline asymmetry in the (charge-symmetric) proton-proton collisions at the LHC,
\begin{eqnarray}\label{eq:incline-asymmetry-qguess-lhc}
A^{\varphi,q} & \equiv & \frac{\sigma_A^{\varphi}(y_{t\bar tj} > 0) - \sigma_A^{\varphi}(y_{t\bar tj} < 0)}{\sigma_S} \approx \frac{\sigma_A^{\varphi}(q\bar q\rightarrow t\bar t j) - \sigma_A^{\varphi}(\bar q q\rightarrow t\bar t j)}{\sigma_S}\,.
\end{eqnarray}
The cross section $\sigma_A^{\varphi}$ is derived from its partonic analog $\hat{\sigma}_A^{\varphi}$ in (\ref{eq:aphi}) by folding it with the PDFs. Since the initial state in the $q\bar q$ channel is antisymmetric under charge conjugation, one has $\sigma_A^{\varphi}(\bar q q\rightarrow t\bar tj) = -\sigma_A^{\varphi}(q \bar q\rightarrow t\bar tj)$. If the identification of the quark direction via $|y_{t\bar t j}|$ was perfect,  $A^{\varphi,q}$ would thus essentially measure the charge asymmetry in the $q\bar q$ channel, $\sigma_A^{\varphi}(q\bar q\rightarrow t\bar t j)/(\sigma_S/2)$. Mistaken $\bar q q$ events with $y_{t\bar t j} < 0$, which appear as $q\bar q$ events, smear out the asymmetry because they contribute to $A^{\varphi,q}$ with the opposite sign. In Figure~\ref{fig:aphi-lhc}, left, we display the misinterpretation rate $R_q^{\text{mis}} = N_{q\bar q}(y_{t\bar tj} < 0,|y_{t\bar tj}|_{\text{min}})/N_{q\bar q}(|y_{t\bar tj}|_{\text{min}})$ (dotted curve) for the LHC at $\sqrt{S}=8\,\text{TeV}$ (LHC8) as a function of a lower cut on the final-state boost, $|y_{t\bar tj}|_{\text{min}}$. Without cuts, the probability to misinterpret a $q\bar q$ state as a $\bar q q$ state amounts to $R_q^{\text{mis}} = 25\,\%$, but can be reduced to below $R_q^{\text{mis}} = 10\,\%$ by focusing on boosted events with $|y_{t\bar tj}|_{\text{min}} \gtrsim 0.9$. Simultaneously, the cut on $|y_{t\bar tj}|$ enhances the partonic initial states with a valence quark over the charge-symmetric $gg$ background. This effect is visualized in Figure~\ref{fig:aphi-lhc}, left, in terms of the fractions $R_{q\bar q} = N_{q\bar q}/N_{\text{tot}}$ and $R_{qg} = N_{qg}/N_{\text{tot}}$ (dashed curves), denoting the number of $q\bar q + \bar q q$ and $qg + gq$ states over the total number of events for a fixed cut $|y_{t\bar tj}|_{\text{min}}$. In order to observe a sizeable incline asymmetry at the LHC, a focus on highly-boosted events thus seems to be indispensable. The price to pay is a significant reduction of the cross section, displayed as $R_{\text{tot}} = N_{\text{tot}}(|y_{t\bar tj}| \ge |y_{t\bar tj}|_{\text{min}})/N_{\text{tot}}$ (plain curve). For instance, a cut of $|y_{t\bar tj}| \ge 0.5$ reduces the cross section to $50\,\%$, a stronger cut of $|y_{t\bar tj}| \ge 1$ implies a reduction to $20\,\%$.

In Figure~\ref{fig:aphi-lhc}, right, the incline asymmetry $A^{\varphi,q}(\theta_j)$ at the LHC8 is shown as a function of the jet scattering angle $\theta_j$ in the parton frame. Without any further cuts (plain curve), the asymmetry is small, reaching a maximum of no more than $A^{\varphi,q}(\theta_j = \pi/2) = -1.5\,\%$. 
The effect of a cut on the final-state boost is visualized for $|y_{t\bar tj}|\ge 1$ (dot-dashed curve). This cut enhances the maximal asymmetry by more than a factor of two, yielding $A^{\varphi,q}(\theta_j = \pi/2, \linebreak|y_{t\bar tj}|\ge1) = -3.3\,\%$. 

The incline asymmetry $A^{\varphi,q}$ is tailored to probe the charge asymmetry at the LHC in the $q\bar q$ channel. However, the $qg$ channel also contributes to the observable and thereby ``pollutes'' the measurement of the $q\bar q$ contribution. The contribution of $qg + gq$ and $\bar q g + g\bar q$ states to $A^{\varphi,q}$ is shown in Figure~\ref{fig:aphi-lhc}, right (dotted curve). At $\theta_j = \pi/2$, where the incline asymmetry reaches its maximum, the $qg$ pollution contributes $+0.09\,\%$ to the asymmetry. 
The effect is even smaller, when the asymmetry is integrated over a range of $\theta_j$ around the maximum. Since this is the preferred region where the observable is collinear-safe, we conclude that for practical purposes the $qg$ pollution can be neglected to a good approximation. The incline asymmetry $A^{\varphi,q}$ therefore provides a largely clean access to the charge asymmetry in the $q\bar q$ channel.

\begin{figure}[!t]
\begin{center}
\includegraphics[height=7.7cm]{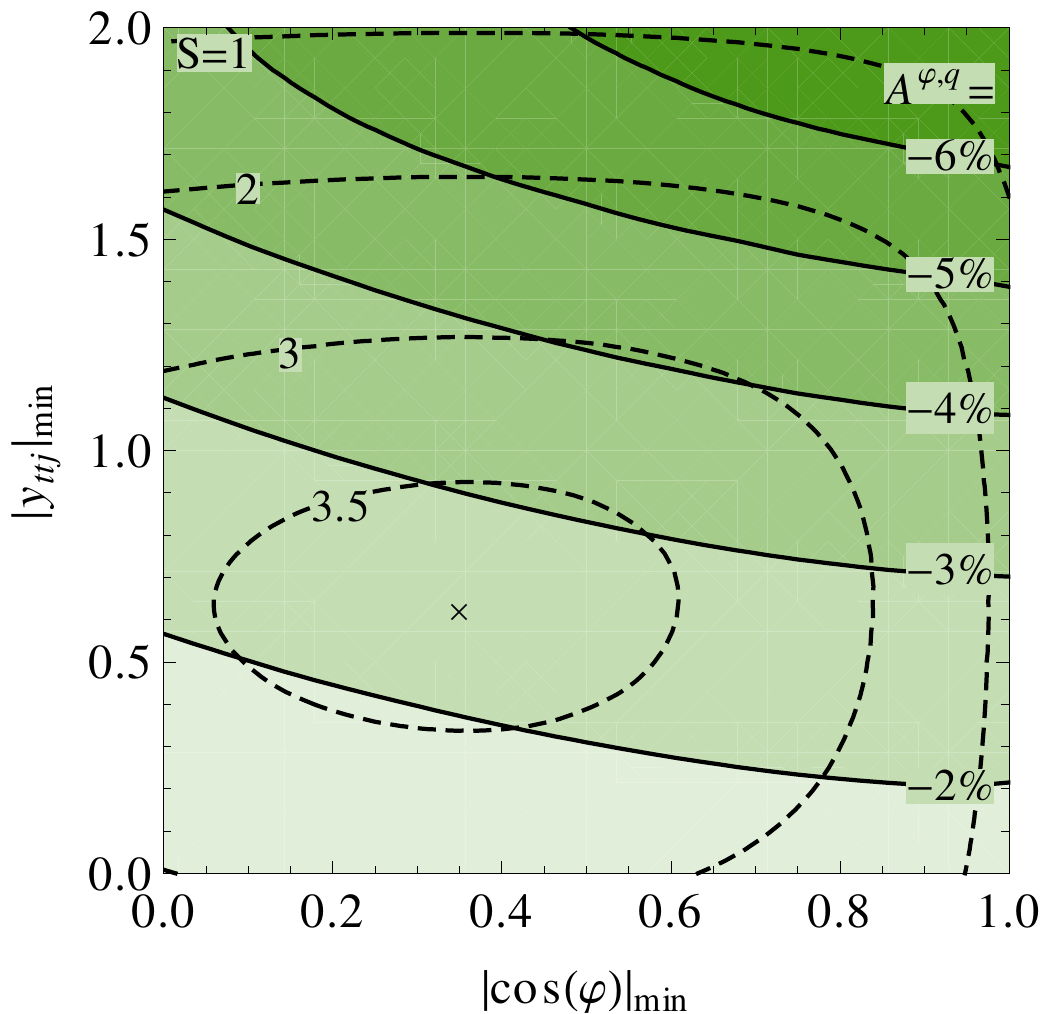}
\hspace*{0.4cm}
\includegraphics[height=7.7cm]{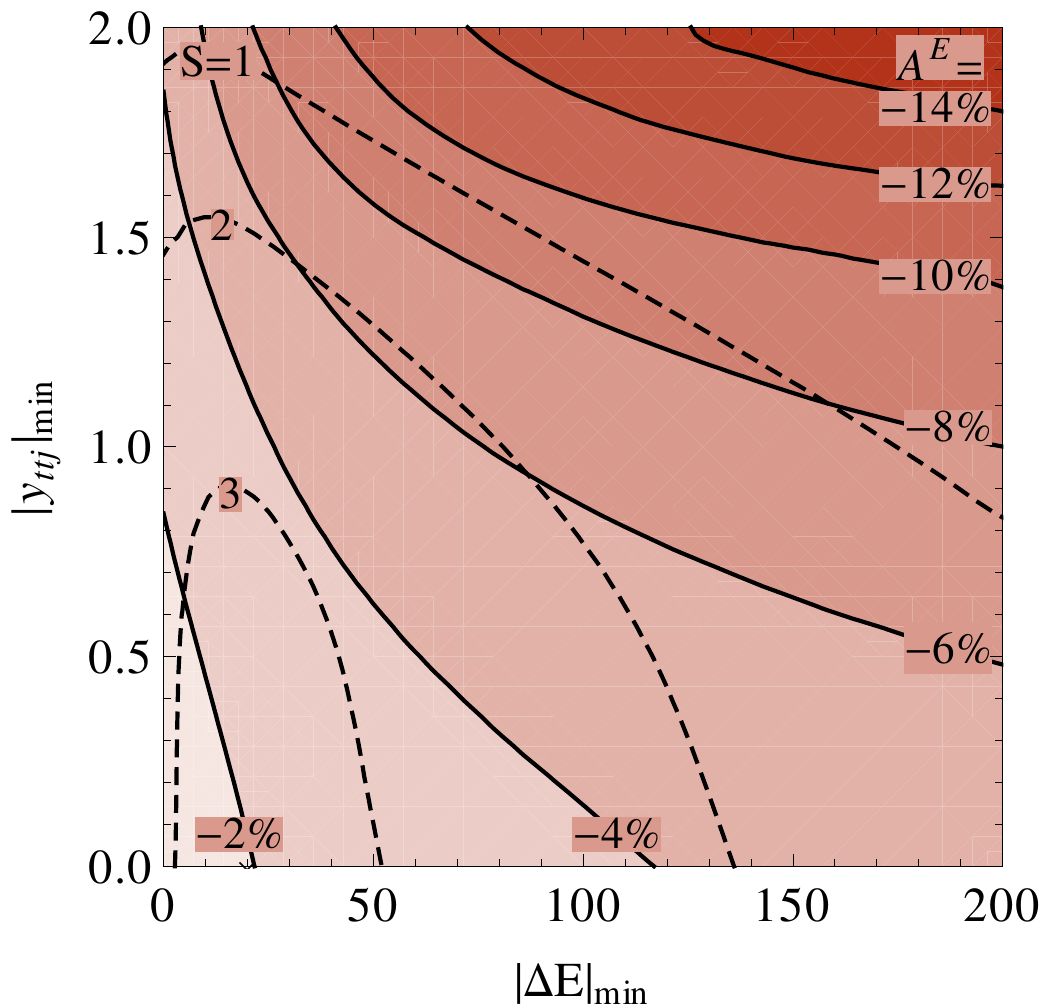}
\end{center}
\vspace*{-1cm}
\begin{center} 
  \parbox{15.5cm}{\caption{\label{fig:aphi-ae-lhc8}Incline asymmetry $A^{\varphi,q}$ (left) and energy asymmetry $A^E$ (right) at LHC8, as functions of the cuts $\{|\cos\varphi|_{\text{min}},|y_{t\bar tj}|_{\text{min}}\}$ and $\{|\Delta E|_{\text{min}},|y_{t\bar tj}|_{\text{min}}\}$, respectively. An additional fixed cut on the partonic jet rapidity, $|\hat{y}_j| \le 1$, has been applied, as well as detector cuts $p_T^j \ge 25\,\text{GeV}$ and $|y_j| \le 2.5$. Superimposed are contour lines of constant asymmetry (plain) and statistical significance $\mathcal{S}=|A|/\delta A$ (dashed).}}
\end{center}
\end{figure}

We now compare the incline asymmetry $A^{\varphi,q}$ with the asymmetry in terms of the top-quark angle $\theta_t$. We determine the quark direction as we did for the incline asymmetry and define the top-angle asymmetry at the LHC as
\begin{eqnarray}\label{eq:afb-lhc}
 A^{\theta_t,q} & \equiv & \frac{\sigma_A^{\theta_t}(y_{t\bar tj} > 0) - \sigma_A^{\theta_t}(y_{t\bar tj} < 0)}{\sigma_S}\,.  
\end{eqnarray}
The observable $A^{\theta_t,q}$ (Figure~\ref{fig:aphi-lhc}, right, dashed curve) meets the incline asymmetry $A^{\varphi,q}$ at its maximum at $\theta_j = \pi/2$ and is consistently smaller for other jet angles. Since the determination of the quark direction does not affect the characteristics of the asymmetries in the partonic CM frame, the incline asymmetry is superior to the top-angle asymmetry at the LHC as well. The jet angular distribution of the asymmetry $A^{\theta_t,q}$ is almost identical to the asymmetry $A^{|y|}$ in terms of absolute top and antitop rapidities (dashed curve),
\begin{eqnarray}\label{eq:ay-lhc}
A^{|y|} & = & \frac{\sigma(\Delta|y| > 0) - \sigma(\Delta|y| < 0)}{\sigma(\Delta|y| > 0) + \sigma(\Delta|y| < 0)}\,,\qquad \Delta|y| = |y_t| - |y_{\bar t}|\,,  
\end{eqnarray}
which has been investigated in inclusive top-pair production at the LHC. The slight {dis\-cre\-pan\-cy} between $A^{|y|}$ and $A^{\theta_t,q}$ is due to the different incorporation of the quark direction and the fact that $A^{|y|}$ is defined in the laboratory frame, whereas $A^{\theta_t,q}$ is defined in the parton frame. The enhancement of the total asymmetry $A^{\varphi,q}$ over $A^{|y|}$ is $50\,\%$ for the applied detector cuts. As for the Tevatron asymmetry, in the region of central jet emission, the enhancement is less pronounced.

\begin{table}[!b]
\begin{center}
\begin{tabular}{|c||c|c|c|c||c|c|c|}
\hline
$|\hat{y}_{j}|_{\text{max}}$ & --- & --- & $1$ & --- & $1$ & $0.5$ & $1$\tabularnewline
$|y_{t\bar tj}|_{\text{min}}$ & --- & $1$ & --- & --- & $0.625$ & $0.7$ & $1.5$ \tabularnewline
$|\cos\varphi|_{\text{min}}$ & --- & --- & --- & $0.8$ & $0.35$ & $0.35$ & $0.7$ \tabularnewline
\hline
\hline
$A^{\varphi,q}\,\,[\%]$ & $-0.84$ & $-2.1$ & $-1.3$ & $-1.1$ & $-2.4$ & $-3.0$ & $-5.0$\tabularnewline
\hline 
$\sigma_{S}\,\,[\text{pb}]$ & $97.5$ & $19.2$ & $51.5$ & $45.5$ & $20.0$ & $9.1$ & $1.8$\tabularnewline
\hline
\hline
$\mathcal{S}(22\,\text{fb}^{-1})$ & $2.8$ & $3.1$ & $3.0$ & $2.5$ & $3.6$ & $3.0$ & $2.3$\tabularnewline  
\hline
\end{tabular}
\end{center}
\begin{center} 
  \parbox{15.5cm}{\caption{\label{tab:lhc8-cuts} Incline asymmetry $A^{\varphi,q}$, cross section $\sigma_S$ and statistical significance $\mathcal{S} = |A^{\varphi,q}|/\delta A^{\varphi,q}$ at LHC8. Detector cuts $p_T^j \ge 25\,\text{GeV}$ and $|y_j| \le 2.5$ have been applied, as well as additional cuts on the boost of the final state, $y_{t\bar tj}$, the jet rapidity in the parton frame, $\hat{y}_j$, and the inclination angle $\varphi$.}}
\end{center}
\end{table}

Due to the large $gg$ background at the LHC, it is crucial to enhance the sensitivity to the charge asymmetry by phase space cuts. For the incline asymmetry, appropriate variables for such cuts are the final-state boost $y_{t\bar tj}$, the partonic jet rapidity $\hat{y}_j$, and the inclination angle $\cos\varphi$: $y_{t\bar tj}$ helps to suppress the $gg$ background, whereas $\hat{y}_j$ and $\cos\varphi$ are useful to enhance the asymmetry at the parton level. Since their effects on the incline asymmetry are to a large extent uncorrelated, cuts on these variables can be adjusted independently in order to maximize the significance of the signal. An upper cut on the jet rapidity $|\hat{y}_j|$ is indispensable to suppress the region of collinear jet emission. We thus preselect a region of central jets and explore the phase space in terms of $\cos\varphi$ and $y_{t\bar tj}$. In Figure~\ref{fig:aphi-ae-lhc8}, left, the incline asymmetry $A^{\varphi,q}$ at the LHC8 is displayed for lower cuts $|\cos\varphi|_{\text{min}}$ and $|y_{t\bar tj}|_{\text{min}}$ and a fixed cut of $|\hat{y}_j| \le 1$. With strong cuts, it is possible to increase the asymmetry to up to $A^{\varphi,q} = -6\,\%$. However, as the luminosity recorded at the LHC8 is limited to $\mathcal{L}=22\,\text{fb}^{-1}$ per experiment, looser cuts are preferred to achieve an optimal significance. We determine the statistical significance as we did for the Tevatron analysis, assuming an experimental efficiency of $5\,\%$. At the point of maximal significance, $\mathcal{S}(22\,\text{fb}^{-1}) = 3.6$ (black cross), the asymmetry amounts to $A^{\varphi,q} = -2.4\,\%$. 

In Table~\ref{tab:lhc8-cuts}, we give numerical values for the incline asymmetry $A^{\varphi,q}$, the cross section $\sigma_S$ and the statistical significance $\mathcal{S}$ with a luminosity of $\mathcal{L} = 22\,\text{fb}^{-1}$ at the LHC8 for different sets of cuts. Since none of the variables is outstanding in increasing $A^{\varphi,q}$, the combination of cuts leads to a better significance than a strong cut on one single variable. The third-to-last and the last column correspond to the regions in Figure~\ref{fig:aphi-ae-lhc8}, left, of maximal significance (black cross) and large asymmetry (upper right corner). The second-to-last column shows the effect of a stronger cut on the partonic jet rapidity. Such a cut increases the asymmetry, but lowers the significance due to the reduction of the cross section. With the LHC8 data set and a statistical significance of three standard deviations, an incline asymmetry of up to $A^{\varphi,q} = -4\,\%$ is expected to be observable. The maximum of the asymmetry, $A^{\varphi,q} = -6\,\%$, however, is difficult to access due to the limited amount of data.

\begin{figure}[!t]
\begin{center}
\includegraphics[height=7.7cm]{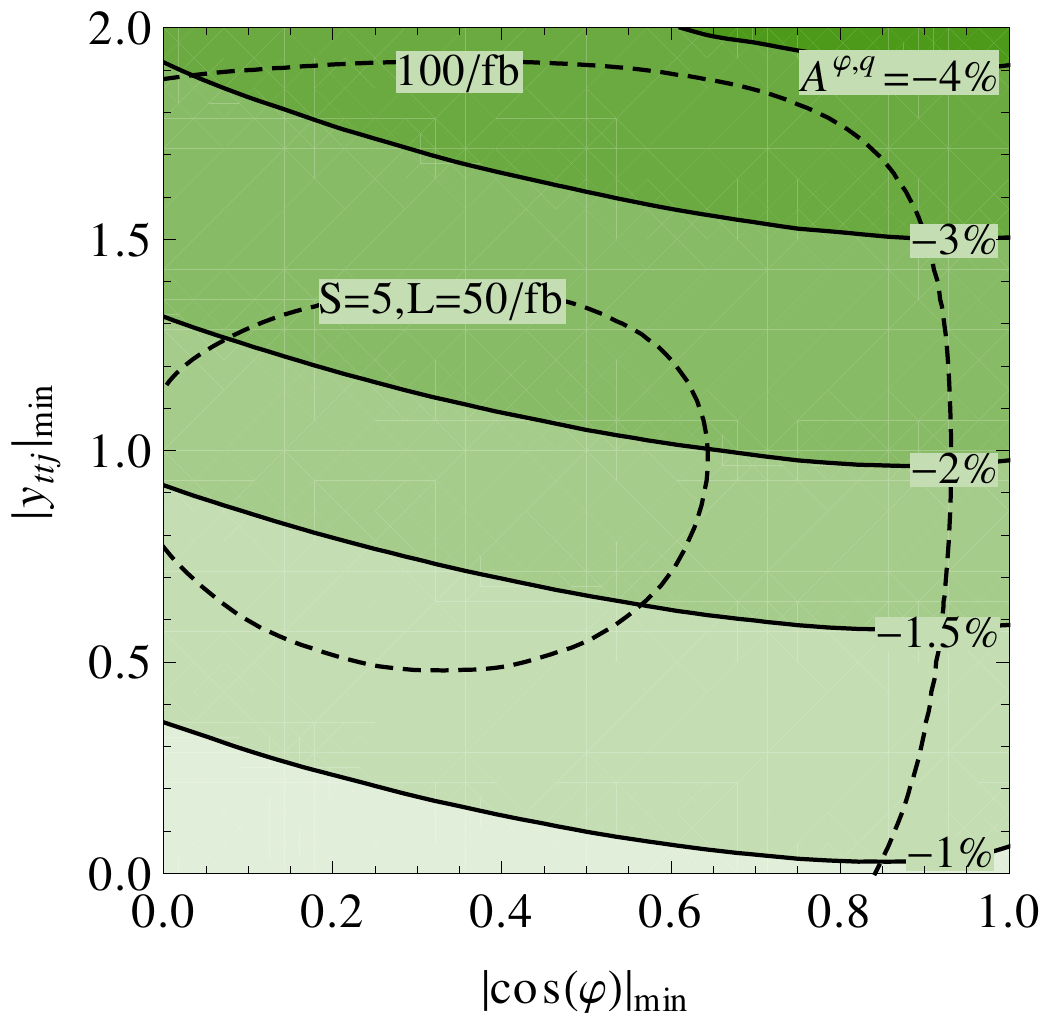}
\hspace*{0.4cm}
\includegraphics[height=7.7cm]{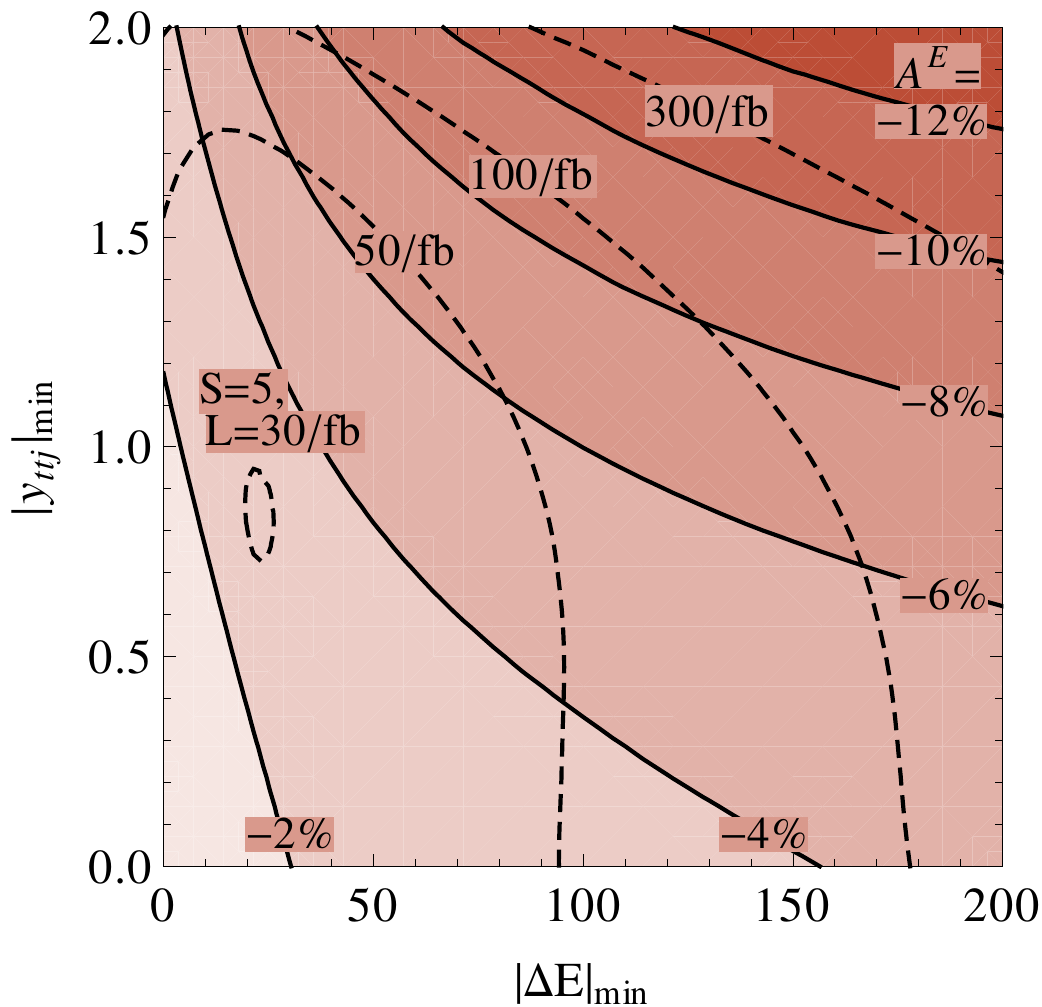}
\end{center}
\vspace*{-1cm}
\begin{center} 
  \parbox{15.5cm}{\caption{\label{fig:aphi-ae-lhc14}Incline asymmetry $A^{\varphi,q}$ (left) and energy asymmetry $A^E$ (right) at LHC14, as functions of the cuts $\{|\cos\varphi|_{\text{min}},|y_{t\bar tj}|_{\text{min}}\}$ and $\{|\Delta E|_{\text{min}},|y_{t\bar tj}|_{\text{min}}\}$, respectively. An additional fixed cut on the partonic jet rapidity, $|\hat{y}_j| \le 0.5$, has been applied, as well as the detector cuts $p_T^j \ge 25\,\text{GeV}$ and $|y_j| \le 2.5$. Superimposed are contour lines of constant asymmetry (plain) and of constant statistical significance $\mathcal{S}=5$ for various luminosities (dashed).}}
\end{center}
\end{figure}

\begin{table}[!b]
\begin{center}
\begin{tabular}{|c||c|c|c|c||c|c|c|}
\hline
$|\hat{y}_{j}|_{\text{max}}$ & --- & --- & $0.5$ & --- & $0.5$  & $0.5$ & $0.25$ \tabularnewline
$|y_{t\bar tj}|_{\text{min}}$ & --- & $1.5$ & --- & --- & $1.3$ & $1.85$ & $1.9$ \tabularnewline
$|\cos\varphi|_{\text{min}}$ & --- & --- & --- & $0.8$ & $0.5$ & $0.7$ & $0.9$ \tabularnewline
\hline
\hline
$A^{\varphi,q}\,\,[\%]$ & $-0.41$ & $-1.4$ & $-0.77$ & $-0.54$ & $-2.4$ & $-3.7$ & $-4.2$\tabularnewline
\hline 
$\sigma_{S}\,\,[\text{pb}]$ & $458$ & $41.3$ & $117$ & $217$ & $17.6$ & $3.6$ & $1.0$\tabularnewline
\hline
\hline
$\mathcal{S}(50\,\text{fb}^{-1})$ & $4.4$ & $4.4$ & $4.2$ & $4.0$ & $5.0$ & $3.5$ & $2.1$\tabularnewline  
\hline
$\mathcal{S}(100\,\text{fb}^{-1})$ & $6.2$ &$6.3$ & $5.9$ & $5.7$ & $7.1$ & $5.0$ & $3.0$\tabularnewline  
\hline
\end{tabular}
\end{center}
\begin{center} 
  \parbox{15.5cm}{\caption{\label{tab:lhc14-cuts} Incline asymmetry $A^{\varphi,q}$, cross section $\sigma_S$ and statistical significance $\mathcal{S} = |A^{\varphi,q}|/\delta A^{\varphi,q}$ at LHC14. Framework as in Table~\ref{tab:lhc8-cuts}.}}
\end{center}
\end{table}

Finally, we comment on the prospects to observe the incline asymmetry at the LHC running at its design collision energy of $\sqrt{S} = 14\,\text{TeV}$ (LHC14). Keeping the same detector cuts as for the LHC8, the total cross section for $t\bar t + j$ production at the LHC14 is $\sigma_S=458\,\text{pb}$ at LO, the partonic contributions amounting 
to $75\%$ ($gg$), $21\%$ ($qg + \bar q g$) and $4\%$ ($q\bar q$).  
Since the $q\bar q$ contribution is almost a factor of $2$ smaller than at the LHC8, 
stronger cuts are required in order to distinguish the incline asymmetry from its $gg$ and $qg$ background. As shown in Figure~\ref{fig:aphi-ae-lhc14}, left, for $|\hat{y}_j| \le 0.5$, the maximal incline asymmetry hardly exceeds $A^{\varphi,q} = -4\,\%$. In turn, the higher luminosity expected at the LHC14 allows us to compensate for this drawback and to achieve a higher statistical significance than at the LHC8. The dashed lines in Figure~\ref{fig:aphi-ae-lhc14}, left, correspond to contours of constant significance $\mathcal{S}=5$ for the luminosities $\mathcal{L} = 50\,\text{fb}^{-1}$ and $\mathcal{L} = 100\,\text{fb}^{-1}$. In Table~\ref{tab:lhc14-cuts}, numerical examples are given for different sets of cuts. The third-to-last column shows the maximal incline asymmetry reachable with a luminosity of $\mathcal{L} = 50\,\text{fb}^{-1}$ and a significance of $\mathcal{S}=5$, $A^{\varphi,q} = -2.4\,\%$. The  second-to-last column shows the maximal incline asymmetry reachable with $\mathcal{L} = 100\,\text{fb}^{-1}$ and $\mathcal{S}=5$, $A^{\varphi,q} = -3.7\,\%$. With $\mathcal{L} = 100\,\text{fb}^{-1}$, an asymmetry of $A^{\varphi,q} = -4.2\,\%$ can be observed with $\mathcal{S}=3$ (last column). These examples show that a good level of statistical significance for the incline asymmetry should be reached within the early runtime of the LHC14.

\subsection{Energy asymmetry at the LHC}\label{subsec:ealhc}
\begin{figure}[!t]
\begin{center}
\includegraphics[height=5.5cm]{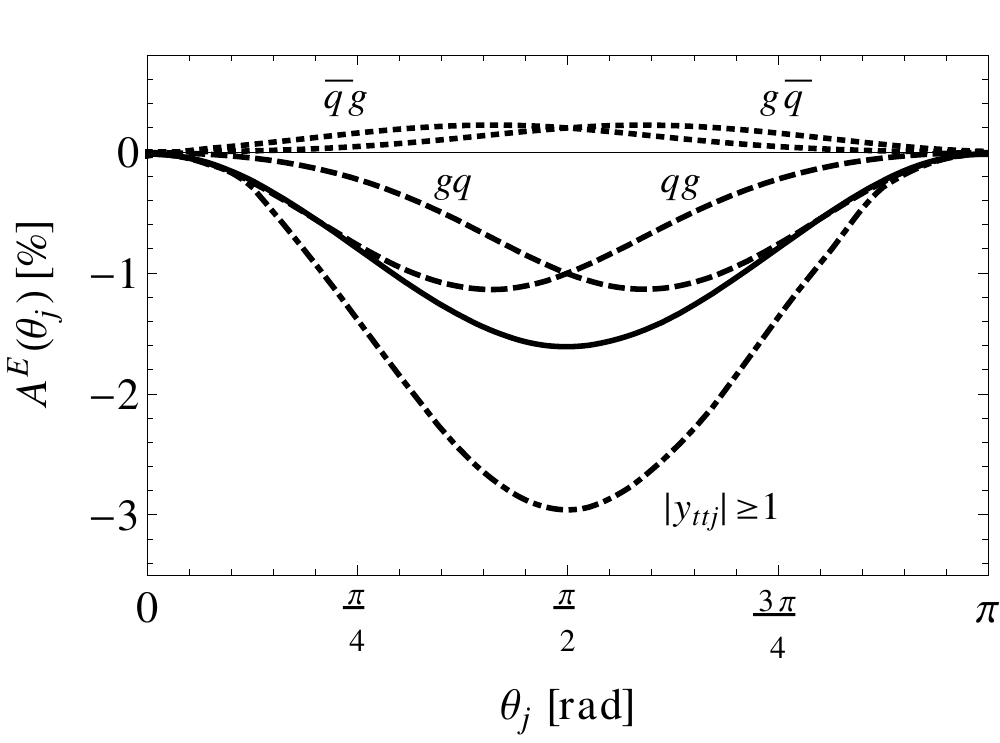}
\hspace*{0.5cm}
\end{center}
\vspace*{-1cm}
\begin{center} 
  \parbox{15.5cm}{\caption{\label{fig:ae-lhc} Energy asymmetry $A^{E}(\theta_j)$ at LHC8 without (plain) and with a boost cut of $|y_{t\bar tj}| \ge 1$ (dot-dashed), as well as the contributions from $qg/gq$ states (dashed) and the pollution from $\bar{q}g/g\bar{q}$ states (dotted). Detector cuts of $p_T^j \ge 25\,\text{GeV}$ and $|y_j| \le 2.5$ have been applied. 
}}
\end{center}
\end{figure}

As has been discussed in Section~\ref{subsec:aenergy}, the energy asymmetry allows us to access the charge asymmetry in the $qg$ channel. At the LHC8, the $qg$ contribution to the total $t\bar t + j$ cross section amounts to $27\,\%$, exceeding the $q\bar q$ contribution by more than a factor of three. The relative abundance of $qg$ states implies less charge-symmetric background for the energy asymmetry and thus facilitates an observation at the LHC. Moreover, unlike the incline asymmetry, the energy asymmetry can be formulated for proton-proton collisions without involving the direction of the incoming quark. We define the energy asymmetry for the LHC as
\begin{eqnarray}\label{eq:energy-asymmetry-lhc}
A^E \,& = &\, \frac{\sigma_A^E}{\sigma_S} \,\, \approx\, \frac{\sigma_A^E(qg\rightarrow t\bar t j) + \sigma_A^E(gq\rightarrow t\bar t j)}{\sigma_S}\,,
\end{eqnarray}
where the hadronic asymmetric cross section $\sigma_A^E = \sigma(\Delta E > 0) - \sigma(\Delta E < 0)$ is obtained from the partonic one in (\ref{eq:energy-asymmetry-qg}) by folding it with the PDFs. Contributions from $q\bar q$ and $\bar q q$ states cancel exactly for any fixed jet angle $\theta_j$, because the charge asymmetry of the initial state implies $\text{d}\sigma_A^E(q\bar q\rightarrow t\bar t j)/\text{d}\theta_j = - \text{d}\sigma_A^E(\bar q q\rightarrow t\bar t j)/\text{d}\theta_j$. The energy asymmetry $A^E$ is thus free from $q\bar q$ contributions. 
It basically measures the charge asymmetry in the $qg$ channel, \linebreak $\sigma_A^E(qg\rightarrow t\bar tj)/(\sigma_S/2)$, when integrated over a symmetric range around $\theta_j = \pi/2$. In Figure~\ref{fig:ae-lhc}, we show the distribution of the energy asymmetry $A^E$ (plain curve) in terms of $\theta_j$. The maximum is obtained for 
central jet emission, $A^E(\theta_j = \pi/2) = - 1.6\,\%$. 
 The separate contributions from $qg$ and $gq$ states (dashed curves) are not symmetric under $\theta_j \leftrightarrow \pi - \theta_j$, due to the asymmetric jet kinematics. Their sum, however, yields a symmetric distribution, reflecting the symmetry of the hadronic initial state. At the parton level, contributions from $\bar{q}g$ and $g\bar{q}$ states (dotted curves) are of the same magnitude, but opposite sign with respect to $qg$ and $gq$ contributions. At the hadron level, due to the lower abundance of antiquarks inside the proton, the $\bar{q}g + g\bar{q}$ contributions are smaller. They reduce the maximum of the asymmetry from $A^E(\theta_j = \pi/2) = -2.0\,\%$ 
 to $A^E(\theta_j = \pi/2) = -1.6\,\%$.

\begin{table}[!b]
\begin{center}
\begin{tabular}{|c||c|c|c|c||c|c|}
\hline
$|\hat{y}_{j}|_{\text{max}}$ & --- & --- & $0.5$ & --- & $1$ & $1$ \tabularnewline
$|y_{t\bar tj}|_{\text{min}}$ & --- & $1$ & --- & --- & ---  & $0.85$ \tabularnewline
$|\Delta E|_{\text{min}}$ [GeV] & --- & --- & --- & $70$ & $20$ & $60$ \tabularnewline
\hline
\hline
$p_{T,\text{min}}^{j}$ [GeV] & $25$ & $25$ & $25$ & $25$ & $25$ & $70$ \tabularnewline
\hline
\hline
$A^{E}\,\,[\%]$ & $-0.79$ & $-1.7$ & $-1.5$ & $-1.3$ & $-1.9$ & $-5.0$ \tabularnewline
\hline 
$\sigma_{S}\,\,[\text{pb}]$ & $97.5$ & $19.2$ & $25.7$ & $23.4$ & $25.9$ & $2.3$ \tabularnewline
\hline
\hline
$\mathcal{S}(22\,\text{fb}^{-1})$ & $2.6$ & $2.4$ & $2.5$ & $2.0$ & $3.3$ & $2.5$ \tabularnewline  
\hline
\end{tabular}
\end{center}
\begin{center} 
  \parbox{15.5cm}{\caption{\label{tab:ae-lhc8-cuts} Energy asymmetry $A^{E}$, cross section $\sigma_S$ and statistical significance $\mathcal{S} = |A^{E}|/\delta A^{E}$ at LHC8. Detector cuts $p_T^j \ge 25\,\text{GeV}$ and $|y_j| \le 2.5$ have been applied, as well as additional cuts on the boost of the final state, $y_{t\bar tj}$, the jet rapidity in the parton frame, $\hat{y}_j$, and the top-antitop energy difference in the parton frame, $\Delta E$. Combined cuts on $|\hat{y}_{j}|$ and $|\Delta E|$ imply a stronger bound on $p_{T,\text{min}}^{j}$ without changing $A^E$ and $\sigma_S$.}}
\end{center}
\end{table}
Similarly to the incline asymmetry, $A^E$ can be enhanced by cuts on the partonic jet rapidity $\hat{y}_j$ and the final-state boost $y_{t\bar t j}$, which project out the central jet region and suppress the charge-symmetric $gg$ background. An additional lower cut on the top-antitop energy difference $\Delta E$ further enhances the asymmetry, as we discussed at the parton level in Section~\ref{subsec:asparton-summary}. In Figure~\ref{fig:aphi-ae-lhc8}, right, the energy asymmetry $A^E$ at the LHC8 is displayed for variable lower cuts $|\Delta E|_{\text{min}}$ and $|y_{t\bar tj}|_{\text{min}}$ and a fixed cut of $|\hat{y}_j| \le 1$. With strong cuts, the asymmetry reaches up to $A^E = -14\,\%$. 
The significance, however, is strongly limited by the amount of data collected in 2012. 
The maximal significance, $\mathcal{S}(22\,\text{fb}^{-1}) = 3.3$, 
is thus reached for the loose cut $|\Delta E|_{\text{min}} = 20\,\text{GeV}$ and no cut on $|y_{t\bar tj}|$. 
The corresponding asymmetry amounts to $A^E = -1.9\,\%$. 
The effects of different cuts on the asymmetry $A^E$, the cross section $\sigma_S$ and the statistical significance $\mathcal{S}$ are displayed in Table~\ref{tab:ae-lhc8-cuts}. The two last columns correspond to the regions in Figure~\ref{fig:aphi-ae-lhc8}, right, of maximal overall significance (lower left corner) and maximal significance for an asymmetry of $A^E = -5\,\%$ (lower left center). Notice again that a lower cut on $\Delta E$ implies a minimum jet energy $E_j \ge E_j^{\text{min}}$ to ensure momentum conservation in the final state. Combined cuts on $|\Delta E|$ and $|\hat{y}_j|$ (namely the jet angle $\theta_j$) correspond to a lower bound on the jet's transverse momentum $p_T^j = E_j \sin\theta_j$, as indicated by the last entry in line four. A stronger cut on $p_T^j$ is experimentally welcome, as it facilitates the identification of the hard jet in the $t\bar t + j$ final state. With the LHC8 data set, the prospects for the energy asymmetry are similar to those for the incline asymmetry: An energy asymmetry of up to $A^E = - 3.7\,\%$ is expected to be observable with a statistical significance of three standard deviations.  

\begin{table}[!b]
\begin{center}
\begin{tabular}{|c||c|c|c|c||c|c|c|}
\hline
$|\hat{y}_{j}|_{\text{max}}$ & --- & --- & $0.2$ & --- & $0.5$ & $0.5$ & $0.5$\tabularnewline
$|y_{t\bar tj}|_{\text{min}}$ & --- & $1.5$ & --- & --- & $1.4$ & $1.7$ & $1.8$ \tabularnewline
$|\Delta E|_{\text{min}}$ [GeV] & --- & --- & --- & $160$ & $61$ & $80$ & $130$ \tabularnewline
\hline
\hline
$p_{T,\text{min}}^{j}$ [GeV] & $25$ & $25$ & $25$ & $25$ & $80$ & $100$ & $160$ \tabularnewline
\hline
\hline
$A^{E}\,\,[\%]$ & $-0.48$ & $-1.5$ & $-1.1$ & $-0.86$ & $-6.5$ & $-8.8$ & $-11$\tabularnewline
\hline 
$\sigma_{S}\,\,[\text{pb}]$ & $458$ & $41.3$ & $46.4$ & $47.6$ & $2.4$ & $0.64$ & $0.14$\tabularnewline
\hline
\hline
$\mathcal{S}(50\,\text{fb}^{-1})$ & $5.2$ & $4.9$ & $3.6$ & $3.0$ & $5.0$ & $3.5$ & $2.0$\tabularnewline
$\mathcal{S}(100\,\text{fb}^{-1})$ & $7.3$ & $6.9$ & $5.1$ & $4.2$ & $7.1$ & $5.0$ & $2.9$\tabularnewline
$\mathcal{S}(300\,\text{fb}^{-1})$ & $12.7$ & $11.9$ & $8.8$ & $7.3$ & $12.3$ & $8.6$ & $5.0$\tabularnewline
\hline
\end{tabular}
\end{center}
\begin{center} 
  \parbox{15.5cm}{\caption{\label{tab:ae-lhc14-cuts} Energy asymmetry $A^{E}$, cross section $\sigma_S$ and statistical significance $\mathcal{S} = |A^{E}|/\delta A^{E}$ at LHC14. Framework as in Table~\ref{tab:ae-lhc8-cuts}. 
}}
\end{center}
\end{table}

At the LHC14, higher luminosities allow us to set stronger cuts on the energy asymmetry and still achieve a good statistical significance. Compared to LHC8, the maximal asymmetry is only slightly reduced due to the larger $gg$ background. In Figure~\ref{fig:aphi-ae-lhc14}, right, we display the energy asymmetry at the LHC14 for $|\hat{y}_j| \le 0.5$. An energy asymmetry of $A^E = -5\,\%$ can be observed with a statistical significance of $\mathcal{S} = 5$ already for luminosities less than $\mathcal{L} = 50\,\text{fb}^{-1}$. With higher luminosities of about $\mathcal{L} = 300\,\text{fb}^{-1}$, it is possible to explore the region of large asymmetries around $A^E = -10\,\%$. From the statistical point of view, the energy asymmetry is thus the preferred observable to probe the charge asymmetry at the LHC14. Compared to the incline asymmetry, it is sizeable for strong cuts and therefore benefits even more from high statistics. In Table~\ref{tab:ae-lhc14-cuts}, numerical examples are given for different sets of cuts. The three last columns show the maximum incline asymmetries for $\mathcal{S}=5$  and $|\hat{y}_j| < 0.5$ obtained with $\mathcal{L} = 50,100,300\,\text{fb}^{-1}$. They amount to $A^{\varphi,q} = -6.5,-8.8,-11\,\%$.  
The implicit constraints on the jet's transverse momentum $p_T^j$ from combined cuts on $|\Delta E|$ and $|\hat{y}_j|$ are shown in line four. Generically, the minimal $p_T^j$ at the LHC14 is higher than at the LHC8, because the preferred cuts on $|\Delta E|$ and $|\hat{y}_j|$ are much stronger. Therefore, a strong detector cut on $p_T^j$ can be set without affecting the magnitude of the cross section and the asymmetry at the LHC14. These examples demonstrate once more the good prospects of observing a sizeable energy asymmetry at the LHC14.

\section{Conclusions}\label{sec:conclusions}
Top-quark pair production in association with a hard jet provides a new way to access the charge asymmetry at hadron colliders. From a theoretical perspective, the charge asymmetry in $t\bar t + j$ production is complementary to the asymmetry in inclusive $t\bar t$ production, as both originate from the interplay of different contributions in QCD. From a phenomenological point of view, the additional jet is advantageous insofar as it allows us to construct new observables beyond the hitherto considered rapidity asymmetries. The discovery potential of the charge asymmetry via these new observables in $t\bar t + j$ production at the LHC is expected to be superior to inclusive $t\bar t$ production. To construct optimal observables, jet kinematics is crucial because it affects the top- and antitop-quark kinematics due to momentum conservation in the final state. Furthermore, jet kinematics governs the cross section in the soft and collinear regime and thereby confines the observability of a charge asymmetry to specific phase space regions.

In this work, we have presented a pair of new observables that respect the jet kinematics, the incline asymmetry and the energy asymmetry. These observables probe the charge asymmetry in $t\bar t + j$ production in both the $q\bar q$ and the $qg$ channel. In the $q\bar q$ channel, the incline and the energy asymmetry are complementary, as they are sensitive to independent parts of the differential charge asymmetry. In the $qg$ channel, this complementarity is lost because the $qg$ initial state is neither symmetric nor antisymmetric under charge conjugation. An incline asymmetry can be defined in the $t\bar t$ rest frame. In practice, however, the energy asymmetry in the $qg$ frame proves very useful. It is equivalent to an angular asymmetry of the top and the antitop quark with respect to the direction of quark-jet and therefore independent of the momentum constellation in the initial state. 

At the Tevatron, the charge asymmetry in $t\bar t + j$ production is accessible via the incline asymmetry $A^{\varphi}$ in the $q\bar q$ channel. Due to the collinear enhancement of the cross section (the symmetric background to the charge asymmetry), the incline asymmetry is maximized if the jet is emitted perpendicular to the beam axis in the $q\bar q$ CM frame. Integrated over the jet angle, the incline asymmetry is sizeable, amounting to $A^{\varphi} = -15.6\,\%$ at LO. It can be increased further by focusing on the region of central jet emission with a cut on the partonic jet rapidity $|\hat{y}_j|$. The experimental sensitivity to the incline asymmetry is limited by the amount of data collected at the Tevatron. Yet, with an estimated efficiency of $5\,\%$ and an integrated luminosity of $10\,\text{fb}^{-1}$, a statistical significance of $4-5$ standard deviations is expected. The energy asymmetry, in turn, is tiny and therefore not observable at the Tevatron.

At the LHC, an observation of the incline asymmetry in the $q\bar q$ channel relies on knowing the direction of the incoming quark. The latter can be estimated by exploiting the boost of the final state, which reflects the large momentum fraction carried by the valence quark inside the proton. The resulting incline asymmetry $A^{\varphi,q}$, however, is strongly suppressed by the large charge-symmetric $gg$ background. By combining cuts on three variables, one can efficiently raise $A^{\varphi,q}$ to an observable level. A cut on the final-state boost $|y_{t\bar tj}|$ enhances the fraction of $q\bar q$ events with respect to the $gg$ background. The collinear region is suppressed as at the Tevatron by selecting events with small jet rapidities $|\hat{y}_j|$. The charge-asymmetric cross section can eventually be enhanced by focusing on small inclination angles $\varphi$. At the LHC8 with the full data set of $22\,\text{fb}^{-1}$, a significance of $3.6$ standard deviations can be obtained for moderate cuts on the three above-mentioned variables, corresponding to $A^{\varphi,q} = -2.4\,\%$. At the LHC14 with an increased luminosity of $100\,\text{fb}^{-1}$, an incline asymmetry of almost $A^{\varphi,q} = -4\,\%$ can be observed with a significance of $5$ standard deviations.

The energy asymmetry is the first observable that allows us to probe the charge asymmetry in the $qg$ channel at the LHC. Exploring the $qg$ channel is particularly useful, because the $gg$ suppression of the asymmetry is much milder than for the $q\bar q$ channel. Importantly, the definition of the energy asymmetry does not require the determination of the quark direction in the initial state. At the LHC8, with suitable cuts on $|y_{t\bar t j}|$, $|\hat{y}_j|$, and on the top-antitop energy difference $\Delta E$, the energy asymmetry reaches a maximum of $A^E\approx - 14\,\%$ at LO. This maximum is twice as large as for the incline asymmetry, reflecting the increased parton luminosity of $qg$ with respect to $q\bar q$ initial states. The limited statistics at LHC8 confine the statistical significance to at most $3.3$ standard deviations, corresponding to an energy asymmetry of $A^E = -1.9\,\%$. At the LHC14, the maximal asymmetry is slightly lowered to $A^E\approx -12\,\%$ due to the increased $gg$ background. An observation of this maximum asymmetry with a significance of $3$ standard deviations requires $100\,\text{fb}^{-1}$ of data. Obviously, the prospects to observe a large energy asymmetry at the LHC14 increase with the accumulation of more data.

Compared to the so far considered rapidity asymmetries, the new incline and energy asymmetries improve the accessibility of the charge asymmetry in $t\bar t + j$ production in two ways. The incline asymmetry can be considered as a refined rapidity asymmetry that takes the jet kinematics into account. Both observables are equal for central jet emission. Integrated over the jet angular distribution, the incline asymmetry is consistently larger than the rapidity asymmetry. The energy asymmetry in addition provides us with access to the hitherto unexplored charge asymmetry in the $qg$ channel. We suggest a measurement of both the incline asymmetry as well as the energy asymmetry in $t\bar t + j$ production at the LHC, in order to obtain complementary results for the charge asymmetry in the $q\bar q$ and the $qg$ channel. With the sizeable energy asymmetry, $t\bar t + j$ production is likely to become the discovery channel of the charge asymmetry at the LHC, given the limited prospects of the rapidity asymmetries in inclusive $t\bar t$ production.

Concerning the numerical predictions in this work, a caveat is in order. Since NLO corrections to the charge asymmetry in $t\bar t + j$ production are important, an analysis beyond LO is necessary to provide robust QCD predictions for the new observables. The theory of $t\bar t + j$ production and top-quark decay at NLO is well understood from investigations of the {ra\-pi\-di\-ty} asymmetries. To what extent the incline and energy asymmetries are affected by NLO corrections is an interesting open question.

\section{Acknowledgments}
We thank Rohini Godbole, Kirill Melnikov, and Markus Schulze for interesting discussions and Hubert Spiesberger for helpful comments on the manuscript. This work was supported by the Initiative and Networking Fund of the Helmholtz Association, contract HA-101 (`Physics at the Terascale'), by the Research Center `Elementary Forces and Mathematical Foundations' of the Johannes Gutenberg-Universit\"at Mainz, and by the National Science Foundation, grant PHY-1212635.


\begin{appendix}

\section{Appendix}\label{app:boosted}
In this appendix, we demonstrate that the energy asymmetry and the top-angle asymmetry in the channels $q\bar q\rightarrow t\bar t g$ and $qg\rightarrow t\bar t q$ are related by a boost into the $t\bar t$ rest frame. This connection is due to the fact that both channels are related by parton crossing $p_2\leftrightarrow \nolinebreak[4]p_3$.\footnote{The partons $p_2$ and $p_3$ are assigned as in Section~\ref{subsec:kinematics}.} The energy asymmetry in either channel is equal to a forward-backward asymmetry of parton $3$ (the jet) in the $t\bar t$ rest frame. The top-angle asymmetry in the respective other channel is approximately equal to the forward-backward asymmetry of parton $2$ (an incoming antiquark or gluon) in the $t\bar t$ rest frame. Based on this relation, it is possible to understand the magnitude of the respective asymmetries. It will become clear why in the $q\bar q$ channel the top-angle asymmetry is so much larger than the energy asymmetry, whereas the situation is reversed in the $qg$ channel.

\begin{figure}[!t]
\begin{center}
\includegraphics[height=5.5cm]{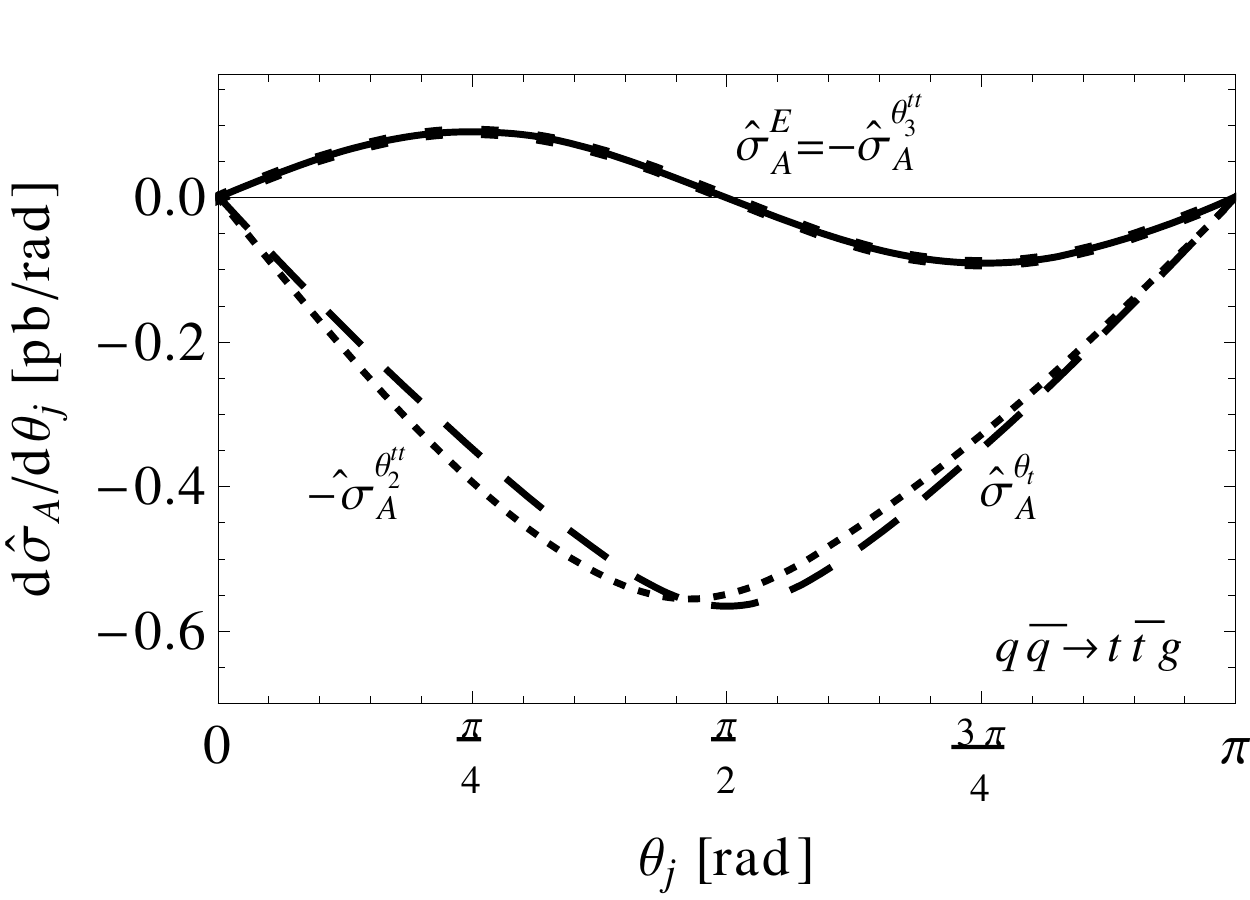}
\hspace*{0.5cm}
\includegraphics[height=5.5cm]{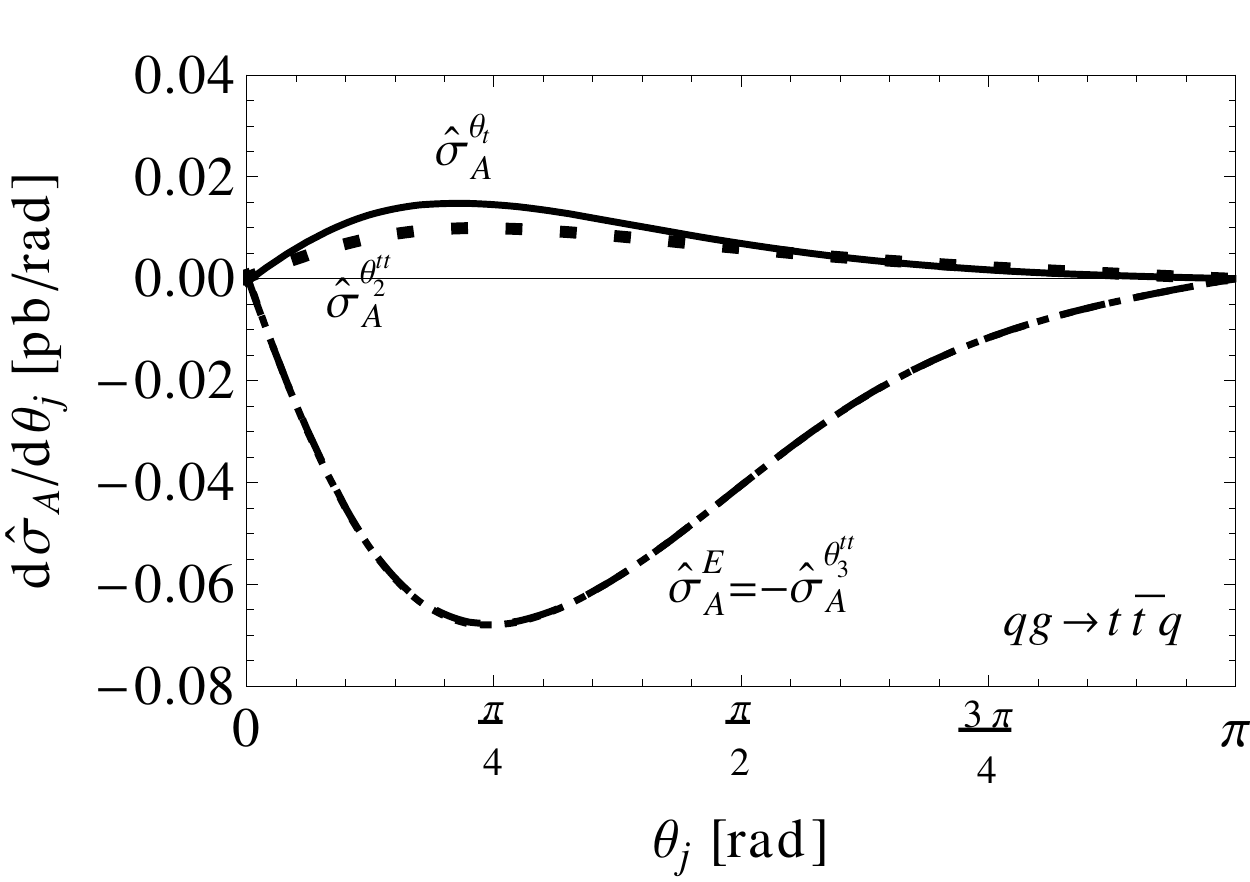}
\end{center}
\vspace*{-1cm}
\begin{center} 
  \parbox{15.5cm}{\caption{\label{fig:a-diff_restframes}Jet angular distributions of partonic asymmetries $\text{d}\hat{\sigma}_A/\text{d}\theta_j$ for $\sqrt{s}=1$\,TeV and $E_j \ge 20$\,GeV in the $q\bar{q}\to t\bar{t} g$ channel (left panel) and the $qg\to t\bar{t} q$ channel (right panel). The jet angle $\theta_j$ is defined in the CM frame of the incoming partons. $\hat{\sigma}_{A}^{\theta_t}$ denotes the top-angle asymmetry, and  $\hat{\sigma}_{A}^E$ the energy asymmetry in the CM system of the incoming partons. $\hat{\sigma}_A^{\theta_i^{t\bar{t}}}$ is the forward-backward asymmetry of the parton $p_i$ in the $t\bar{t}$ rest frame.
}}
\end{center}
\end{figure}

We start with an observation made in Section~\ref{subsec:aenergy}: The energy asymmetry $\hat{\sigma}_A^E$ as defined in (\ref{eq:sigmaa-energy}) is (up to a sign) equivalent to an angular asymmetry $\hat{\sigma}_A^{\xi}$ of the top and antitop quarks with respect to the jet direction, defined in (\ref{eq:energy-angular}). We can also express $\hat{\sigma}_A^{\xi}$ as a rapidity asymmetry with respect to the jet momentum $\vec{k}_3$,
\begin{eqnarray}
\hat{\sigma}_A^{y^{k_3}} & \equiv & \hat{\sigma}(\Delta y^{k_3} > 0) - \hat{\sigma}(\Delta y^{k_3} < 0) = - \hat{\sigma}_A^E\,,\qquad \Delta y^{k_3} = y_t^{k_3} - y_{\bar{t}}^{k_3}\,,
\end{eqnarray}
where the top and antitop rapidities with respect to $\vec{k}_3$ are given by
\begin{eqnarray}
 y_t^{k_3} & = & \frac{1}{2} \ln \frac{E_t+ \cos \xi \cdot |\vec{k}_t|}{E_t - \cos \xi \cdot |\vec{k}_t|}
\,, \qquad 
 y_{\bar{t}}^{k_3} \,\, = \,\, \frac{1}{2} \ln \frac{E_{\bar{t}}+ \cos \bar{\xi} \cdot |\vec{k}_{\bar{t}}|}{E_{\bar{t}} - \cos \bar{\xi} \cdot |\vec{k}_{\bar{t}}|}\,.
\end{eqnarray}%
Since the rapidity difference $\Delta y^{k_3}$ is invariant under boosts along $\vec{k}_3$ and since $\vec{k}_t+\vec{k}_{\bar{t}}+\vec{k}_3=\vec{0}$, it is convenient to boost the system into the $t\bar t$ rest frame, where $\vec{k}'_t = -\vec{k}_{\bar t}'$. In the $t\bar t$ rest frame, the rapidity asymmetry $\hat{\sigma}_A^{y^{k_3}}$ appears as a forward-backward asymmetry of the jet with respect to the top-quark direction,
\begin{eqnarray}\label{eq:fb-as-ttb}
\hat{\sigma}_A^{\theta_3^{t\bar t}} & \equiv & \hat{\sigma}(\cos\theta_3^{t\bar t} > 0) - \hat{\sigma}(\cos\theta_3^{t\bar t} < 0)\,,\qquad
\end{eqnarray}
where $\theta_3^{t\bar t}$ denotes the scattering angle of parton $3$ with respect to the top-quark momentum in the $t\bar t$ rest frame. The energy asymmetry $\hat{\sigma}_A^E$ is thus (up to a sign) equal to the forward-backward asymmetry of the jet in the $t\bar t$ rest frame, $\hat{\sigma}_A^{\theta_3^{t\bar t}}$. We illustrate this result in Figure~\ref{fig:a-diff_restframes}, where we display the jet angular distributions of the asymmetries for the $q\bar q$ channel (left) and the $qg$ channel (right). The energy asymmetry $\hat{\sigma}_A^E$ in the partonic CM frame (left, plain / right, dashed) and the forward-backward asymmetry $-\hat{\sigma}_A^{\theta_3^{t\bar t}}$ in the $t\bar t$ rest frame (left, bold dotted / right, fine dotted) coincide. The relation $\hat{\sigma}_A^E = -\hat{\sigma}_A^{\theta_3^{t\bar t}}$  holds in both the $q\bar q$ and the $qg$ channel, since it is independent from the partonic origin of the jet, but only based on kinematic relations between the final-state momenta. Whether the energy asymmetry in the partonic CM frame or the forward-backward asymmetry in the $t\bar t$ rest frame is more appropriate for a measurement remains to be examined on an experimental level.

We will now establish the relation between the energy asymmetry $\hat{\sigma}_A^E$ and the top-angle asymmetry $\hat{\sigma}_A^{\theta_t}$ in different parton channels by using the forward-backward asymmetry in the $t\bar t$ rest frame. The top-angle asymmetry, as defined in (\ref{eq:top-angle-as}) and (\ref{eq:top-angle-distribution}), can be considered as the asymmetry of the top- and antitop-quark rapidities with respect to the incident parton $p_2$. Due to the invariance under boosts along $\vec{k}_2$, this rapidity asymmetry is closely related to the forward-backward asymmetry of parton $2$ in the $t\bar t$ rest frame, $\hat{\sigma}_A^{\theta_2^{t\bar t}}$. The asymmetry $\hat{\sigma}_A^{\theta_2^{t\bar t}}$ is defined as in (\ref{eq:fb-as-ttb}) by changing $\theta_3^{t\bar t}\rightarrow \theta_2^{t\bar t}$. The difference between $\hat{\sigma}_A^{\theta_t}$ and $\hat{\sigma}_A^{\theta_2^{t\bar t}}$ is due to the imbalance of the top and antitop's transverse momenta with respect to the beam axis, caused by the presence of the jet in the final state. Disregarding this small difference, we can eventually express the energy asymmetry in one channel and the top-angle asymmetry in the other channel in terms of forward-backward asymmetries in the $t\bar t$ rest frame, which are related by parton crossing,
\begin{eqnarray}\label{eq:qqb-qg-correspondence}
\hat{\sigma}_A^E(q\bar q) & = & - \hat{\sigma}_A^{\theta_3^{t\bar t}}(q\bar q)\quad \stackrel{p_2\leftrightarrow p_3}{\longleftrightarrow}\quad \ \ \hat{\sigma}_A^{\theta_2^{t\bar t}}(qg)\approx \hat{\sigma}_A^{\theta_t}(qg)\,,\\\nonumber
\hat{\sigma}_A^{\theta_t}(q\bar q) & \approx & -\hat{\sigma}_A^{\theta_2^{t\bar t}}(q\bar q)\quad \stackrel{p_2\leftrightarrow p_3}{\longleftrightarrow}\quad -\hat{\sigma}_A^{\theta_3^{t\bar t}}(qg) = \hat{\sigma}_A^E(qg)\,.
\end{eqnarray}
These results are visualized in Figure~\ref{fig:a-diff_restframes}. The energy asymmetry $\hat{\sigma}_A^E$ in the $q\bar q$ channel (left, plain) and the top-angle asymmetry $\hat{\sigma}_A^{\theta_t}$ in the $qg$ channel (right, plain) can both be related to the forward-backward asymmetry of the gluon in the $t\bar t$ rest frame (bold dotted). The smallness of the top-angle asymmetry in the $qg$ channel has been mentioned previously for inclusive top-pair production in the literature. It can now be ascribed to the small energy asymmetry in the $q\bar q$ channel, since both asymmetries are related by parton crossing. The different shapes of the jet angular distributions are due to the kinematic configurations: In the $q\bar q$ channel, $\theta_j$ is the angle of the gluon-jet with respect to the incident quark, whereas in the $qg$ channel, it is the angle of the quark-jet. In particular, the energy asymmetry in the $q\bar q$ channel is antisymmetric and changes its sign at $\theta_j = \pi/2$. This is not the case for the top-angle asymmetry in the $qg$ channel, where the angle between the incident quark and the gluon is fixed to $\pi$. The magnitudes of $\hat{\sigma}_A^E(q\bar q)$ and $\hat{\sigma}_A^{\theta_t}(qg)$ differ mainly because of the soft enhancement for $E_j\rightarrow 0$ in the $q\bar q$ channel, which is absent in the $qg$ channel.

The same connection holds true for the energy asymmetry in the $qg$ channel (right, dashed) and the top-angle asymmetry in the $q\bar q$ channel (left, dashed). Both asymmetries can be traced back to the forward-backward asymmetry of the quark in the $t\bar t$ rest frame (fine dotted). Thus, the sizeable energy asymmetry in the $qg$ frame is due to the fact that the top-angle asymmetry in the $q\bar q$ channel (which is connected to the incline asymmetry) provides the dominant contribution to the charge asymmetry. These findings confirm that the incline asymmetry in the $q\bar q$ channel and the energy asymmetry in the $qg$ channel capture most of the differential charge asymmetry in both channels.

One might wonder, why we derived the correspondence between the energy asymmetry and the top-angle asymmetry, rather than the incline asymmetry. As was mentioned at the end of Section~\ref{subsec:kinematics}, the incline asymmetry and the energy asymmetry form a ``perfect match'', in the sense that they probe complementary parts of the charge asymmetry in the $q\bar q$ channel. It is not possible to translate this complementarity to the $qg$ channel. A Lorentz boost into the $p_1 p_3$ rest frame, which would mimic a $q\bar q$ initial state, is kinematically prohibited. The incline asymmetry can be defined in the $t\bar t$ rest frame via $\hat{\sigma}_A^{\varphi_{t\bar t}} = \hat{\sigma}(\cos\varphi_{t\bar t} > 0) - \hat{\sigma}(\cos\varphi_{t\bar t} < 0)$, where $\varphi_{t\bar t}$ is the inclination angle between the plane spanned by the top quark and the incoming gluon and the plane spanned by the quark-jet and the incoming gluon. Due to the asymmetric jet kinematics in the $qg$ channel, the jet distribution of $\hat{\sigma}_A^{\varphi_{t\bar t}}$ is not flat as in the $q\bar q$ channel, but peaks in the forward direction. The magnitude of $\hat{\sigma}_A^{\varphi_{t\bar t}}$ is similar to the energy asymmetry $\hat{\sigma}_A^E$ in the $qg$ channel. We consider the energy asymmetry a more suitable observable than the incline asymmetry in the $t\bar t$ rest frame, because the former does not rely on the collinear region. Furthermore, the energy asymmetry can be easily formulated in the partonic CM frame and does not rely on the direction of the incoming gluon, contrary to the incline asymmetry in the $t\bar t$ rest frame.

\end{appendix}

\end{document}